\documentclass[twocolumn,prd,superscriptaddress,preprintnumbers,nofootinbib]{revtex4}[11pt]
\pdfoutput=1
\usepackage{amsmath,amssymb,graphicx} 
\graphicspath{{figs/}}
\usepackage{epsf,verbatim}
\usepackage{hyperref}
\usepackage{comment}
\usepackage{color}
\usepackage{slashed}
\usepackage{subfigure}
\usepackage[usenames,dvipsnames]{xcolor}
\usepackage{comment}
\usepackage{mdwlist, paralist}

\def\to{\rightarrow}

\newcommand{\simgt}{\mathrel{\lower2.5pt\vbox{\lineskip=0pt\baselineskip=0pt
           \hbox{$>$}\hbox{$\sim$}}}}
\newcommand{\simlt}{\mathrel{\lower2.5pt\vbox{\lineskip=0pt\baselineskip=0pt
           \hbox{$<$}\hbox{$\sim$}}}}

\definecolor{jeremycolour}{rgb}{0.2,0.5,0.6}

\newcommand{\squishlist}{
 \begin{list}{$\bullet$}
  { \setlength{\itemsep}{0pt}
     \setlength{\parsep}{3pt}
     \setlength{\topsep}{3pt}
     \setlength{\partopsep}{0pt}
     \setlength{\leftmargin}{1.5em}
     \setlength{\labelwidth}{1em}
     \setlength{\labelsep}{0.5em} } }

\newcommand{\squishlisttwo}{
 \begin{list}{$\bullet$}
  { \setlength{\itemsep}{0pt}
     \setlength{\parsep}{0pt}
    \setlength{\topsep}{0pt}
    \setlength{\partopsep}{0pt}
    \setlength{\leftmargin}{2em}
    \setlength{\labelwidth}{1.5em}
    \setlength{\labelsep}{0.5em} } }

\newcommand{\squishend}{
  \end{list}  }

\newcommand{\uhid}{$U(1)'$}

\newcommand{\be}{\begin{equation}}
\newcommand{\ee}{\end{equation}}
\newcommand{\bea}{\begin{eqnarray}}
\newcommand{\eea}{\end{eqnarray}}

\newcommand{\f}{\frac}

\newcommand{\g}{\text{ GeV}}
\newcommand{\m}{\text{ MeV}}

\def\babar{\mbox{\sl B\hspace{-0.4em} {\small\sl A}\hspace{-0.37em} \sl B\hspace{-0.4em} {\small\sl A\hspace{-0.02em}R}}}

\begin{document}

\title{Constraining Light Dark Matter with Low-Energy $e^+e^-$ Colliders}

\preprint{YITP-SB-23-13}

\author{Rouven Essig}
\thanks{rouven.essig@stonybrook.edu}
\affiliation{C.N.~Yang Institute for Theoretical Physics, Stony Brook University, Stony Brook, NY 11794}

\author{Jeremy Mardon}
\thanks{jmardon@stanford.edu}
\affiliation{Stanford Institute for Theoretical Physics, Department of Physics, Stanford University, Stanford, CA 94305}

\author{Michele Papucci}
\thanks{mpapucci@umich.edu}
\affiliation{Michigan Center for Theoretical Physics, 
University of Michigan, Ann Arbor, MI 48109}
 
\author{Tomer Volansky}
\thanks{tomerv@post.tau.ac.il}
\affiliation{Raymond and Beverly Sackler School of Physics and Astronomy, Tel-Aviv University, Tel-Aviv 69978, Israel}

\author{Yi-Ming Zhong}
\thanks{yiming.zhong@stonybrook.edu}
\affiliation{C.N.~Yang Institute for Theoretical Physics, Stony Brook University, Stony Brook, NY 11794}

%%%%%%%%%%%%%%%%%%%%%%%%%%%%%%%%%%%%%%%%%%%%%%%%%%%
\begin{abstract} 
We investigate the power of low-energy, high-luminosity electron--positron colliders to probe hidden sectors with a mass below 
$\sim 10$~GeV that couple to Standard Model particles through a light mediator.
Such sectors provide well-motivated dark matter candidates, and can give rise to distinctive mono-photon signals at 
$B$-factories and similar experiments. 
We use data from an existing mono-photon search by \babar\ to place new constraints on this class of models, and give projections for the sensitivity of a similar search at a future $B$-factory such as Belle II. 
We find that the sensitivity of such searches are more powerful than searches at other collider or fixed-target facilities for 
hidden-sector mediators and particles with masses between a few hundred MeV and 10~GeV.
Mediators produced on-shell and decaying invisibly to hidden-sector particles such as dark matter can be probed particularly well. 
Sensitivity to light dark matter produced through an off-shell mediator is more limited, but may be improved with a better  control of backgrounds, allowing background estimation and a search for kinematic edges.
We compare our results to existing and future direct detection experiments and show that low-energy colliders  provide an indispensable and complementary avenue to search for light dark matter. 
The implementation of a mono-photon trigger at Belle II would provide an unparalleled window into such light hidden sectors.  
\end{abstract}
%%%%%%%%%%%%%%%%%%%%%%%%%%%%%%%%%%%%%%%%%%%%%%%%%%%

\maketitle

 \setcounter{equation}{0} \setcounter{footnote}{0}

%%%%%%%%%%%%%%%%%%%%%%%%%%%%%%%%%%%%%%%%%%%%%%%%%%%
\section{INTRODUCTION}
\label{sec:intro}
%%%%%%%%%%%%%%%%%%%%%%%%%%%%%%%%%%%%%%%%%%%%%%%%%%%

The search for the identity of dark matter (DM), which makes up 85\% of the matter in our Universe, 
is one of the most important experimental endeavors of our time.  
All evidence for DM comes from its gravitational interactions with ordinary matter.  
However, the success of laboratory and space-based experiments searching for DM is predicated on DM having 
additional non-gravitational interactions with ordinary matter.
A Weakly Interacting Massive Particle (WIMP) is a theoretically well-motivated DM candidate with mass in the 10~GeV to 10~TeV range,
typically interacting with Standard Model (SM) particles through the Electroweak sector.
This makes the WIMP hypothesis testable at many ongoing and upcoming colliders and direct and indirect detection experiments.  
However, there are also many well motivated non-WIMP DM candidates that can also be probed by such experiments.  
In this paper we focus on {\em light} DM (LDM), with mass below 10~GeV, 
interacting with the SM through a light mediator. 
We will show that, among collider experiments,  low-energy $e^+e^-$ colliders such as $B$-factories and $\Phi$-factories are 
particularly well suited to exploring this mass range.  

DM searches at colliders have received much attention in the past.
However, most of the focus has been on searches with high-energy colliders such as LEP, the Tevatron, the LHC, and an ILC, 
see \textit{e.g.}~\cite{Goodman:2010ku,Beltran:2010ww,Bai:2010hh,Goodman:2010yf,Fox:2011fx, Rajaraman:2011wf,Fox:2011pm,Fox:2012ee, Chatrchyan:2012tea, Aad:2012fw,An:2012ue,Fox:2012ru,Zhou:2013fla, Baltz:2006fm, Dreiner:2006sb, Dreiner:2007vm,Dreiner:2009ic}.
These colliders are ideally suited for probing Weak-scale DM, and for DM whose interactions with ordinary matter are mediated by heavy particles.  
In contrast, $B$-factories ($\Phi$-factories) operate at much lower center-of-mass energies of $\sqrt{s} \approx 10$~GeV (1~GeV). 
Their sensitivity is therefore highest to LDM with low-mass mediators.

\begin{figure}[t]
\includegraphics[width=0.22\textwidth]{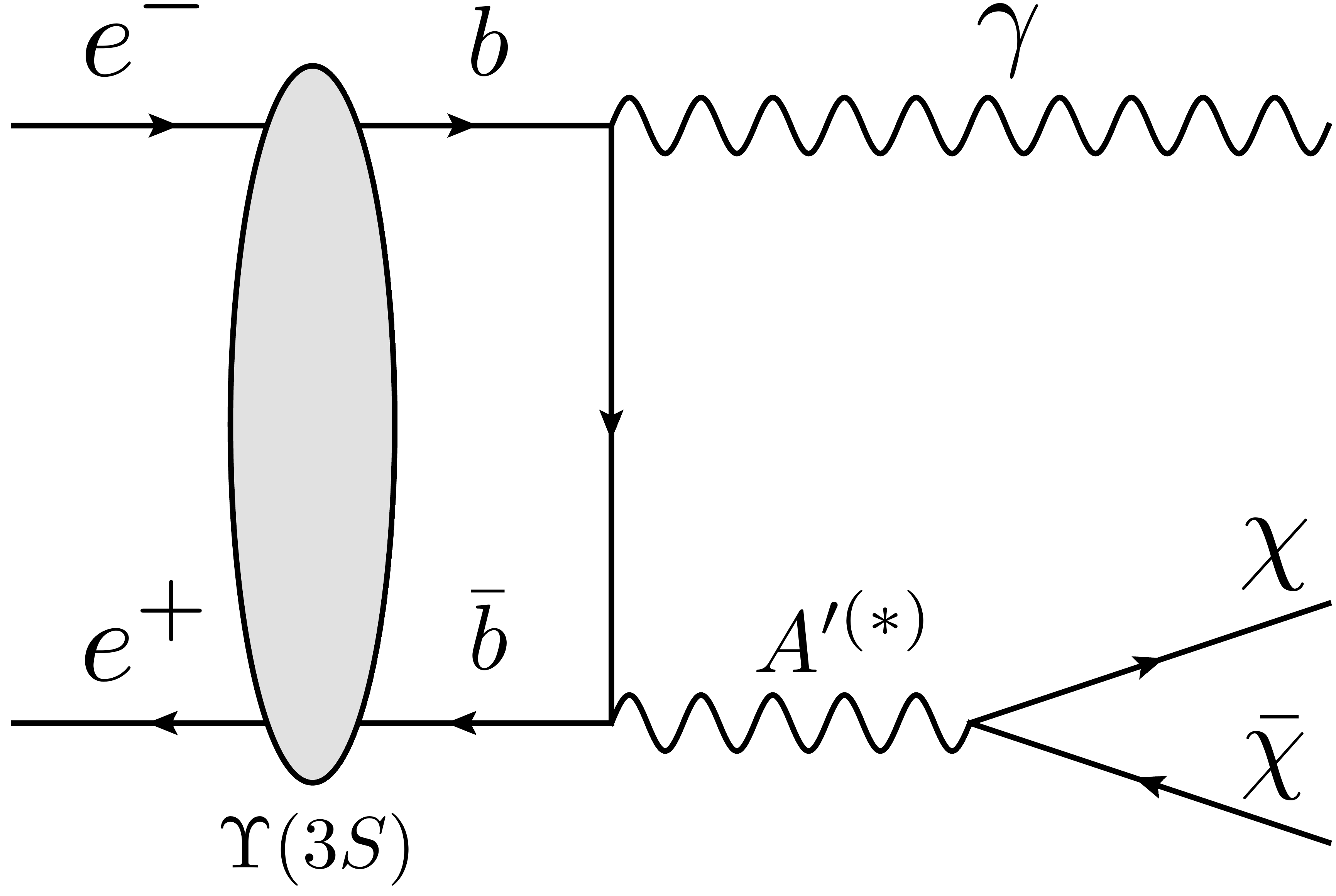}
\includegraphics[width=0.22\textwidth]{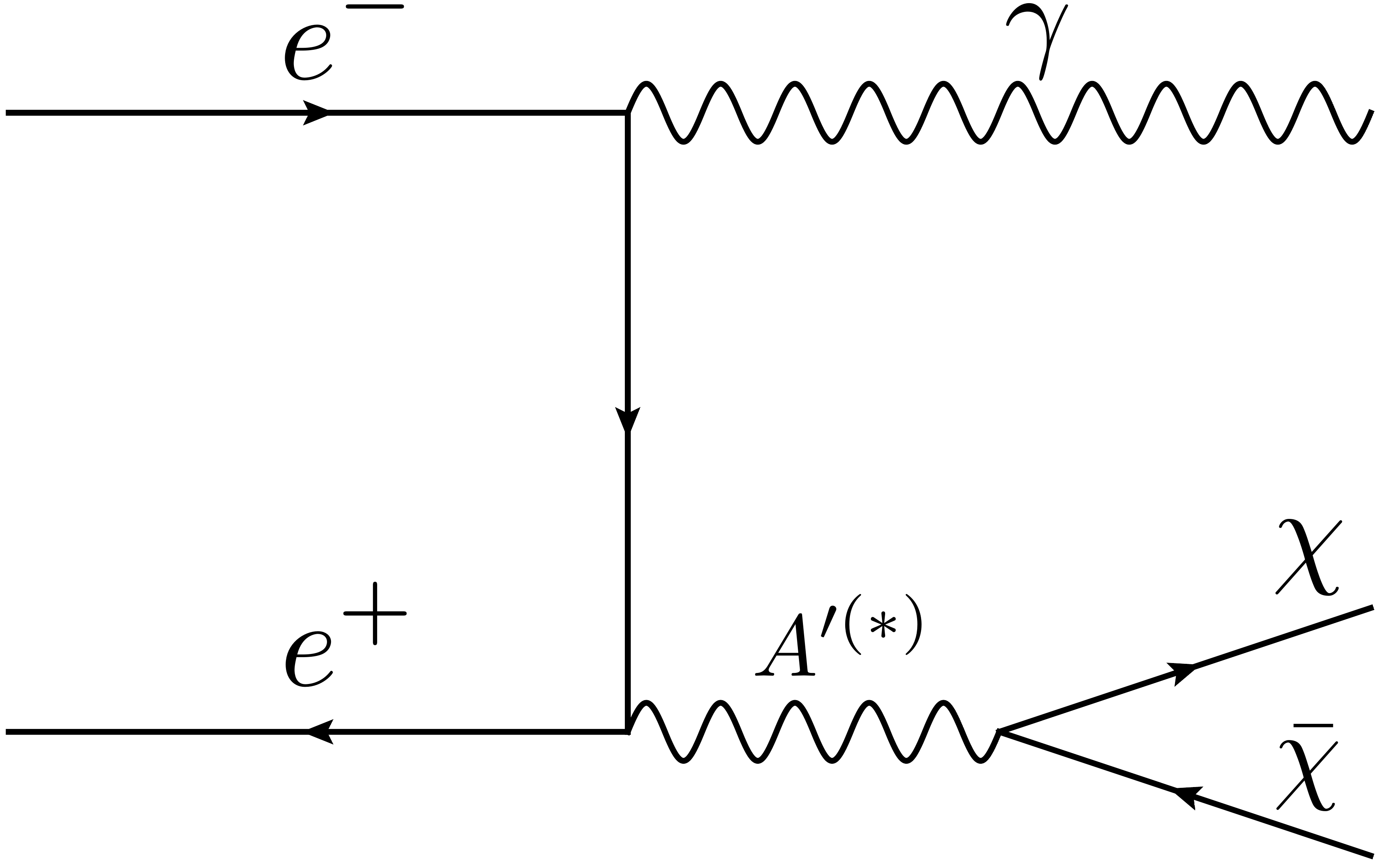}
\caption{
$\gamma + \slashed E$ production channels for LDM coupled through a light mediator.
{\bf Left:} Resonant $\Upsilon(3S)$ production, followed by decay to $\gamma + \chi \, \overline \chi$ through an on- or off-shell mediator.
{\bf Right:} The focus of this paper -- non-resonant $\gamma + \chi \, \overline \chi$ production in $e^+ e^-$ collisions, through an on- or off-shell 
light mediator $A'^{(*)}$.  (Note that in this paper, the symbol $A'$ is used for vector, pseudo-vector, scalar, and pseudo-scalar mediators.) 
}
\label{fig:production}
\end{figure}

DM particles produced in colliders do not scatter in the detector, and appear as missing energy, $\slashed E$.  
A particularly clean channel to study is LDM produced in association with a single photon, 
resulting in a mono-photon signature ($\gamma + \slashed E$).
In this paper, we study the sensitivity of mono-photon searches at low-energy $e^+ e^-$ colliders to LDM.
While LDM production has been studied before in the context of rare meson decays, {\em e.g.}~\cite{Bird:2004ts, McElrath:2005bp, Fayet:2006sp, Bird:2006jd,Rubin:2006gc,Tajima:2006nc, Kahn:2007ru, Dorokhov:2007bd, Fayet:2007ua,Fayet:2009tv,Yeghiyan:2009xc, Dorokhov:2009xs, delAmoSanchez:2010ac, Badin:2010uh,Echenard:2012iq,Calderini:2012ar},  here we consider a complementary possibility: non-resonant production of mono-photon events directly in the $e^+ e^-$ collisions, see 
Fig.~\ref{fig:production} 
(this possibility has previously been considered in less detail in 
e.g.~\cite{Borodatchenkova:2005ct, Fayet:2007ua, Essig:2009nc,Reece:2009un,Dreiner:2009ic}).
We first reanalyse the results of an existing search by the \babar\ collaboration for mono-photon events in decays of the $\Upsilon(3S)$~\cite{Aubert:2008as}.
While \babar\ had an active mono-photon trigger for only $\sim 55$/fb (including $\sim 30$/fb on the $\Upsilon(3S)$) 
out of a total of $\sim 500$/fb of data collected over its lifetime, and performed only a very limited background estimate
on these events, the resulting bounds on LDM improve significantly upon existing bounds in parts of the LDM parameter space.
A similar analysis with Belle or KLOE data is not possible, due to the lack of a single-photon trigger.  

We also estimate the possible sensitivity of Belle II to LDM.  
This will depend strongly on the ability to implement a mono-photon trigger, and to reduce or subtract backgrounds, but should reach substantially beyond the constraints from \babar\ in parts of parameter space.
Our results stress the importance for Belle II to { {\em include a mono-photon trigger during the entire course of data taking}}.  

The rest of the paper is organized as follows. 
In Sec.~\ref{sec:LDM-model-overview} we give a brief theoretical overview of LDM coupled through a light mediator.
Sec.~\ref{sec:production} contains a more detailed discussion of the production of such LDM at low-energy $e^+e^-$ colliders. 
In Sec.~\ref{sec:BaBar} we describe the \babar\ search~\cite{Aubert:2008as}, and extend the results to place constraints on LDM.  
In Sec.~\ref{sec:comparison-to-LEP-DD} we compare our results to existing constraints such as LEP, rare decays, beam-dump experiments, and direct detection experiments. In Sec.~\ref{sec:Belle-II-projections} we estimate the reach of a similar search in a future $e^+ e^-$ collider such as Belle II.
We conclude in Sec.~\ref{sec:conclusion}.
A short appendix discusses the constraints on invisibly decaying hidden photons for some additional scenarios.

%%%%%%%%%%%%%%%%%%%%%%%%%%%%%%%%%%%%%%%%%%%%%%%%%%%
\section{Light Dark Matter with a Light Mediator}
\label{sec:LDM-model-overview}
%%%%%%%%%%%%%%%%%%%%%%%%%%%%%%%%%%%%%%%%%%%%%%%%%%%

A LDM particle, in a hidden sector that couples weakly to ordinary matter through a light, neutral boson (the mediator), is part of many well-motivated frameworks that have received significant theoretical and experimental attention in recent years, 
see 
e.g.~\cite{Boehm:2003hm,Boehm:2003ha,Strassler:2006im,Hooper:2008im,ArkaniHamed:2008qn,Pospelov:2008jd,Pospelov:2007mp, 
Cholis:2008vb, Essig:2010ye,Ruderman:2009tj, Falkowski:2010cm, Morrissey:2009ur, Nomura:2008ru, Baumgart:2009tn, 
Feng:2008ya, Abel:2008ai, Jaeckel:2010ni, Cohen:2010kn} and references therein.  A light mediator may play a significant role in setting the DM relic density~\cite{Lin:2011gj, MarchRussell:2012hi}, or in alleviating possible problems with small-scale structure in $\Lambda$CDM cosmology~\cite{Loeb:2010gj,Tulin:2013teo}.

The hidden sector may generally contain a multitude of states with complicated interactions among themselves.
However, for the context of this paper, it is sufficient to characterize it by a simple model with just two particles, 
the DM particle $\chi$ and the mediator $A'$ (which, with abuse of notation, may refer to a generic (pseudo-)vector, or (pseudo-)scalar, and does not necessarily 
indicate a hidden photon), and four parameters: 
\begin{itemize}
 \item[(i)] $m_\chi$  (the DM mass) 
 \item[(ii)] $m_{A^\prime}$ (the mediator mass)
 \item[(iii)] $g_e$ (the coupling of the mediator to electrons)
 \item[(iv)] $g_\chi$ (the coupling of the mediator to DM).
\end{itemize}
In most of the parameter space only restricted combinations of these four parameters are relevant for $\chi \overline \chi$ production in $e^+ e^-$ collisions; we describe this in more detail in Sec.~\ref{sec:production}.
The spin and $C P$ properties of the mediator and DM particles also have a (very) limited effect on their production rates, but will have a more significant effect on comparisons to other experimental constraints, as will the couplings of the mediator to other SM particles. For the rest of the paper, the ``dark matter'' particle, $\chi$, can be taken to represent any hidden-sector state that couples to the mediator and is invisible in detectors; in particular, it does not have to be a (dominant) component of the DM. 

The simplest example of such a setup is DM that does not interact with the 
SM forces, but that nevertheless has interactions with ordinary matter through a {\em hidden photon}. 
In this scenario, the $A'$ is the massive mediator of a broken Abelian gauge group, \uhid, in the hidden sector, and has a small kinetic mixing, $\varepsilon/\cos{\theta_W}$, with SM hypercharge, $U(1)_Y$~\cite{Holdom:1985ag, Dienes:1996zr, Pospelov:2007mp, ArkaniHamed:2008qn, Pospelov:2008jd, Galison:1983pa, Lin:2011gj}.  %Others?%
SM fermions with charge $q_i$ couple to the $A'$ with coupling strength $g_e = \varepsilon \, e \, q_i$. 
The variables $\varepsilon$, $g_\chi$, $m_\chi$, and $m_{A'}$ are the free parameters of the model.
We restrict 
\bea
g_\chi < \sqrt{4\pi}\,, \;\;\;\;\;\; \mbox{(perturbativity)}
\eea 
in order to guarantee calculability of the model. Such a constraint is also equivalent to imposing $\Gamma_{A'}/m_{A'}\lesssim 1$ which is necessary for the $A'$ to have a particle description. We will refer in the following to this restriction as the ``perturbativity'' constraint. 

In this paper, we discuss this prototype model as well as more general LDM models with vector, pseudo-vector, scalar, and pseudo-scalar mediators. 
We stress that in UV complete models, scalar and pseudo-scalar mediators generically couple to SM fermions through mixing with a Higgs boson, and consequently their coupling to electrons is proportional to the electron Yukawa, $g_e\propto y_e \sim 3 \times 10^{-6}$.
As a result, low-energy $e^+ e^-$ colliders are realistically unlikely to be sensitive to them.  Nonetheless, since more intricate scalar sectors may allow for significantly larger couplings, we include them for completeness. 

For simplicity we consider only fermionic LDM, as the differences between fermion and scalar production are very minor.
We do not consider models with a t-channel mediator (such as light neutralino production through selectron exchange). 
In these, the mediator would be electrically charged and so could not be light.

%%%%%%%%%%%%%%%%%%%%%%%%%%%%%%%%%%%%%%%%%%%%%%%%%%%
\section{Production of Light Dark Matter at $e^+e^-$ Colliders}
\label{sec:production}
%%%%%%%%%%%%%%%%%%%%%%%%%%%%%%%%%%%%%%%%%%%%%%%%%%%

Fig.~\ref{fig:production} illustrates the production of $\gamma + \slashed E$ events at low-energy $e^+e^-$ colliders in LDM scenarios.
The channel shown on the left of  Fig.~\ref{fig:production} is the resonant production of a heavy meson such as $\Upsilon(3S)$, followed by its decay to $\gamma + \chi \overline \chi$ through an on- or off-shell mediator. This channel probes the couplings of the mediator to the $b$-quark (specifically its pseudo-vector or pseudo-scalar couplings if the mediator is on shell).
The focus of this paper, however, is a complementary channel, shown on the right of Fig.~\ref{fig:production}, where LDM is produced through an on- or off-shell mediator, which couples directly to electrons.
 
The collider signal consists of mono-photon events, $\gamma + \slashed E$. The photon energy spectrum can vary quite significantly depending on the masses of the DM and mediator, and we divide the $m_\chi$--$m_{A'}$ plane into three regions with distinct kinematics, as illustrated in Fig.~\ref{fig:search-regions}.
Typical spectra for each region are shown in Fig.~\ref{fig:typicalsim} as a function of the $\chi \overline \chi$ invariant mass, which is related to the photon energy by
\begin{equation}
m^2_{\chi \bar \chi}\equiv s-2 \sqrt{s} E_\gamma^* \, .
\label{eq:invmass}
\end{equation}
The regions are as follows:

\begin{figure}[t]
\begin{center}
%\fbox{
\includegraphics[width=0.35 \textwidth]{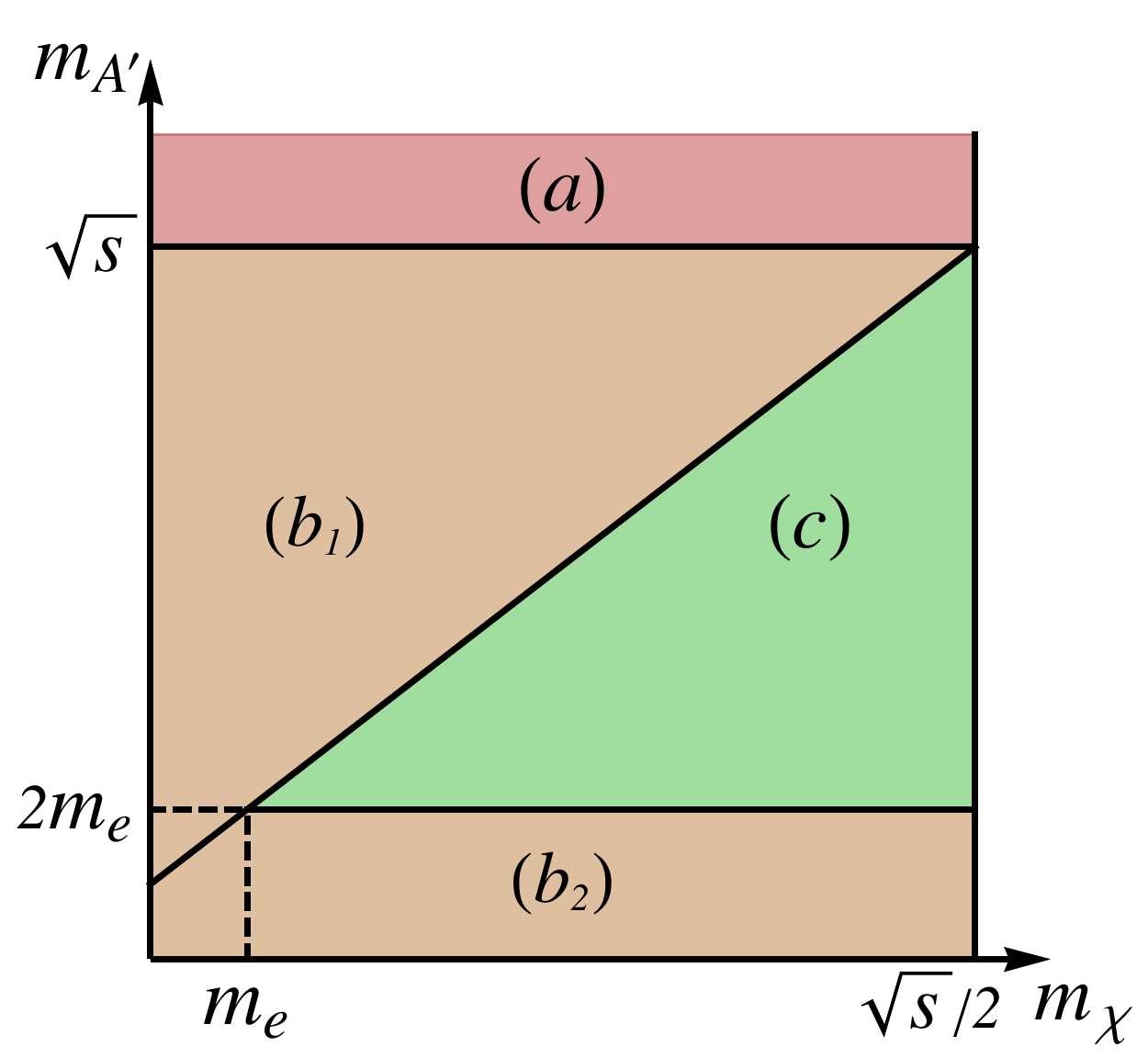}
%}
%\end{center}
\caption{Regions in the $m_{\chi}$--$m_{A'}$ plane with different characteristic $\gamma$+$\slashed E$ signals ($A'$ is any type of mediator). 
{\bf Region~(a)} corresponds to an off-shell heavy mediator, for which an effective operator analysis holds if $m_{A'} \gg \sqrt{s}$.
In {\bf Region~(b)} the mediator is invisible and is produced on-shell. 
In {\bf Region~(c)}, while the mediator is light enough to be produced on shell, $\chi \overline \chi$ production occurs through an off-shell mediator. 
}
\label{fig:search-regions}
\end{center}
\end{figure}

\begin{itemize}

\item[(a):]
$\mathbf{m_{A'} > \sqrt s}$.   Here the mediator is too heavy to be produced on shell, and $\chi \overline \chi$ production proceeds through an off-shell mediator. In this case, the photon is dominantly produced in the form of initial state radiation, and so has a spectrum rising towards low energies (high $m_{\chi \overline \chi}$), illustrated by the red histogram in Fig.~\ref{fig:typicalsim}.

\item[(b):]
$\mathbf{\sqrt s > m_{A'} > 2 m_\chi}$ or $\mathbf{m_{A'} < 2 m_e}$. In region (b$_1$), 
the mediator decays to $\chi \overline \chi$ (the branching fraction to SM particles is assumed to be negligible).
In region (b$_2$), it is too light to decay to either $\chi \overline \chi$ or $e^+ e^-$; if {\em e.g.}~$A'$ is a hidden photon, 
it eventually decays to (three) photons far outside the detector.  
Both cases result in a mono-energetic $\gamma + \slashed E$ signature. 
The spectrum is peaked at $m^2_{\chi \bar \chi} = m_{A'}^2$, but the finite detector resolution gives a width of 
\begin{equation}
\sigma_{m_{\chi \overline \chi}^2} = 2 \sqrt s \, \sigma_{E_\gamma} = (s - m_{\chi \overline \chi}^2) \times (\sigma_{E_\gamma} / E_\gamma) \,,
\label{eq:on-shell-width}
\end{equation}
where $\sigma_{E_\gamma}$ is the experimental photon energy resolution. 
This is illustrated by the two orange histograms in Fig.~\ref{fig:typicalsim}. 

\item[(c):]
$\mathbf{2 m_\chi > m_{A'} > 2 m_e}$.
In this region, the mediator can be produced on shell, but is too light to decay to $\chi\bar\chi$.  
It could either decay to another light hidden-sector state (if available) or it will instead decay to SM fermions, 
an interesting signature, which is not the topic of this paper (see \emph{e.g.}~\cite{Borodatchenkova:2005ct, Fayet:2007ua, Essig:2009nc,Reece:2009un}).    
In the latter case, while direct searches for the \emph{visible} decay are likely to be more sensitive 
(see e.g.~\cite{Bjorken:2009mm,Hewett:2012ns} and references therein), 
$\gamma + \slashed E$ events can occur in the production of $\chi\bar\chi$ through an off-shell mediator. 
Probing these decays is necessary to assess whether such light mediator couples to invisible particles such as LDM and therefore is complementary to visible searches. 
The mediator propagator contributes a factor $1/m_{\chi \overline \chi}^4$ to the cross-section, resulting in the broad, flat photon spectrum illustrated by the green histogram in  Fig.~\ref{fig:typicalsim}.
 
\end{itemize}

\begin{figure*}[ht!]
\centering
\includegraphics[width=0.98\textwidth]{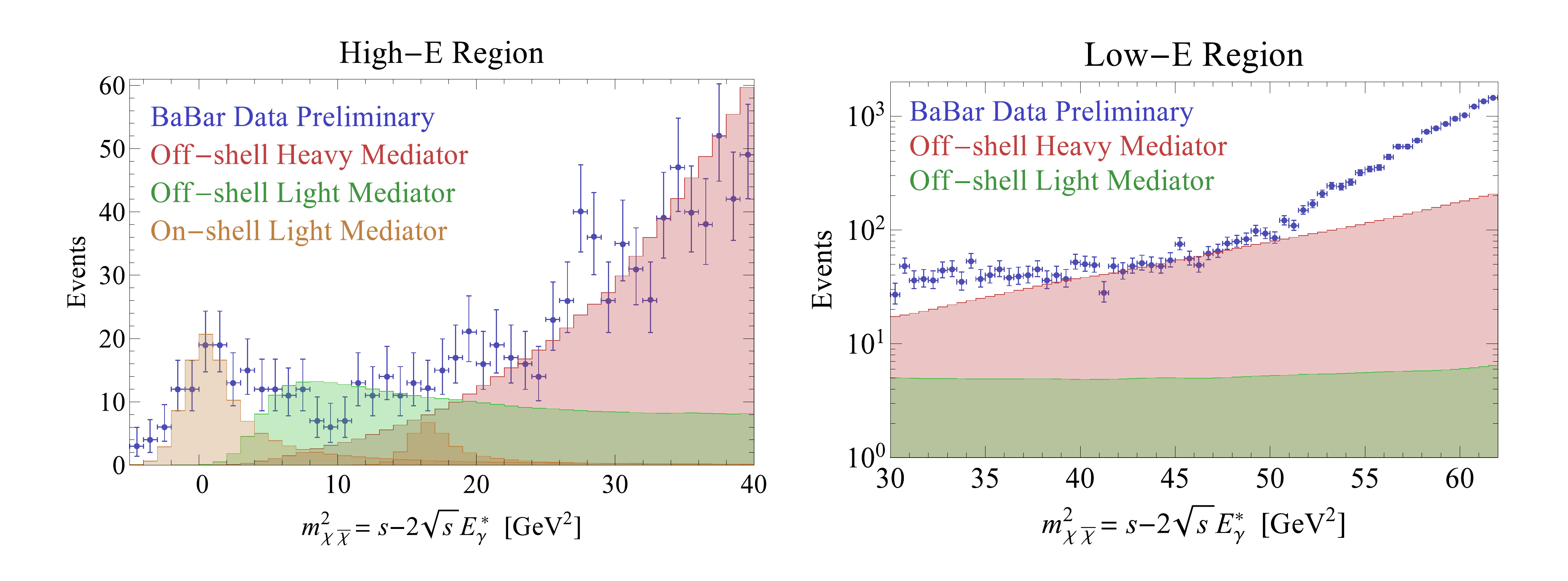}
\vspace{-4mm}
\caption{
Typical simulations of $\gamma$+$\slashed E$ signals compared to data that was scanned from 
the \babar\ Collaboration (unpublished)~\cite{Aubert:2008as}, in both the ``High-E'' ({\bf left}) and ``Low-E'' ({\bf right}) search 
regions (where $3.2 \g<E_\gamma^*<5.5 \g$ and $2.2 \g<E_\gamma^*<3.7 \g$, respectively; see Sec.~\ref{sec:BaBar} for more details).
The {\bf red} histogram illustrates $\chi \overline \chi$ production through an off-shell heavy mediator  (region (a)), resulting in a rising spectrum. 
The histogram corresponds to $m_\chi=1 \g$ and  $m_{A'}=12 \g$. 
The {\bf orange} histograms show the peaked spectra arising from on-shell production of an invisible mediator (region (b)), with $m_{A'}=0.5 \m$ (left) or $4 \g$ (right).  
The {\bf green} histogram shows the typical broad
spectrum resulting from  $\chi \overline \chi$ production through an off-shell light mediator (region (c)) (we show $m_\chi=m_{A'}=1 \g$). 
In each case the cross-section is scaled to lie at the 95\% CL limits presented in Sec.~\ref{sec:BaBar}.
}
\label{fig:typicalsim}
\end{figure*}

\subsection{Relevant parameters}

In each of the three regions, only limited combinations of the four model parameters presented in Sec.~\ref{sec:LDM-model-overview} determine the $\gamma + \slashed E$ signal, with the remaining combinations being redundant (or giving small corrections), as follows:

\begin{itemize}

\item[(a):] 
When $m_{A'} \gg \sqrt s$, the mediator can be integrated out of the theory and the interaction described by a 4-point vertex. For fermionic LDM coupling through a vector, pseudo-vector, scalar or pseudo-scalar mediator, the effective operator describing the interaction is given by (respectively) 
\begin{eqnarray}
\label{eq:eff1}
{\mathcal O}_V &=& \frac{1}{\Lambda^2}\left(\bar\chi \gamma_\mu  \chi \right)\left(\bar e \gamma^\mu e \right)\,,
\\
\label{eq:eff2}
{\mathcal O}_{A} &=& \frac{1}{\Lambda^2}\left(\bar\chi \gamma_\mu \gamma^5  \chi \right)\left(\bar e \gamma^\mu \gamma^5 e \right)\,,
\\
\label{eq:eff3}
{\mathcal O}_S &=& \frac{1}{\Lambda^2}\left(\bar\chi   \chi \right)\left(\bar e  e \right)\,,
\\
\label{eq:eff4}
{\mathcal O}_{PS} &=& \frac{1}{\Lambda^2}\left(\bar\chi  \gamma^5  \chi \right)\left(\bar e  \gamma^5 e \right) \, .
\end{eqnarray}
where $\Lambda$ is given by
\begin{equation}
\Lambda \equiv \frac{m_{A'}}{\sqrt{g_e  g_\chi}} \, .
\label{eq:cutoff}
\end{equation}
The signal spectrum depends on $m_\chi$, and the rate is proportional to $\Lambda^{-4}$,
with corrections of order $m_{\chi \overline \chi}^2/m_{A'}^2$, relevant only for $A'$ masses close to the center-of-mass energy.

\item[(b):]
For mediators produced on shell, $m_\chi$ and $g_\chi$ are irrelevant as long as the mediator does not have a significant branching fraction to SM fermions. 
The signal spectrum is controlled by $m_{A'}$, and the rate is proportional to $g_e^2$,
with corrections of order $g_e^2/g_\chi^2$.

\item[(c):]
For $m_{A'} \ll m_\chi$, the signal spectrum depends on $m_\chi$ but not on $m_{A'}$, 
and the rate is proportional to  $(g_e g_\chi)^2$,
with corrections of order $m_{A'}^2/m_{\chi \overline \chi}^2$.
\end{itemize}

\begin{figure*}[ht!]
\centering
\includegraphics[width=0.96\textwidth]{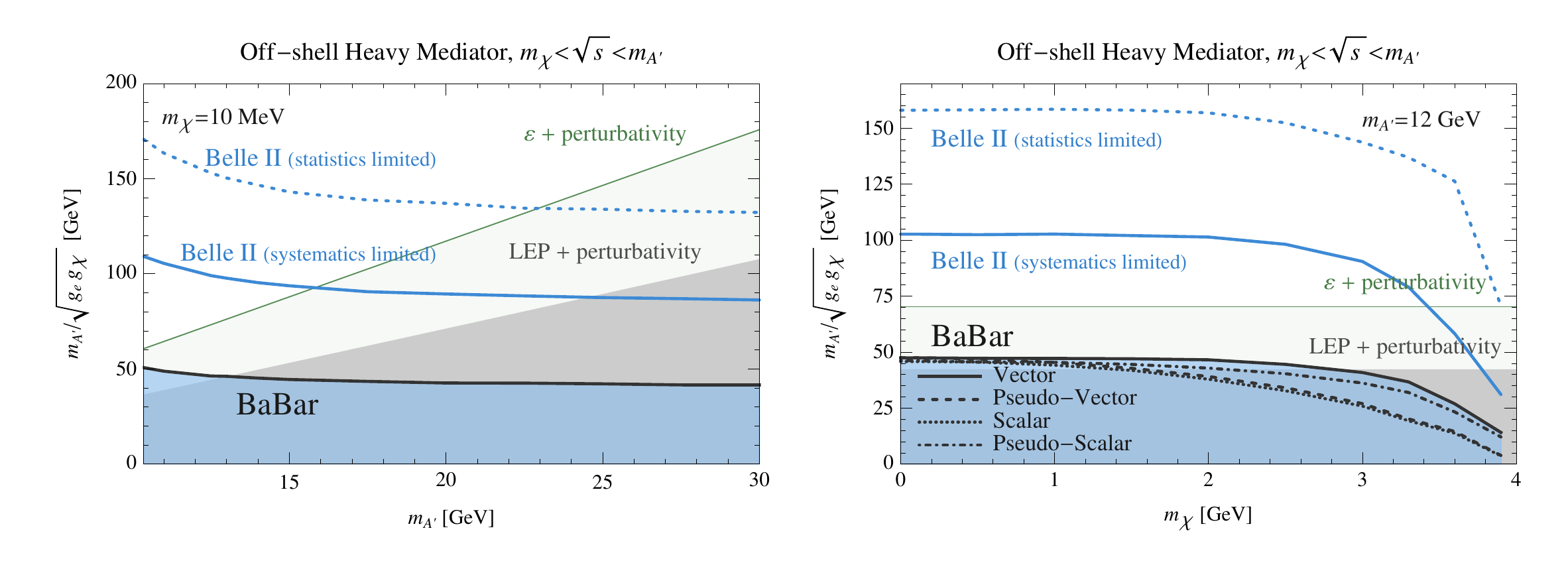}
\vspace{-4mm}
\caption{
Lower bounds on $m_{A'}/\sqrt{g_e  g_\chi}$
in region (a) of Fig.~\ref{fig:search-regions} (production of $\chi \overline \chi$ through a heavy off-shell mediator), for
(\textit{{\bf left}}) a fixed DM mass of 10~MeV, and (\textit{{\bf right}}) a fixed mediator mass of 12~GeV. 
The solid black line / blue shaded region show the bounds from \babar\ data (this work) with a vector mediator. 
On the right, the bounds with other mediators are shown with different 
 line styles, 
while on the left they are almost identical to the vector case and thus not shown separately. 
The solid and dotted blue line both show the projected reach of Belle~II in the vector-mediated case assuming 
that the various background components are known at the $5-20\%$ level (``systematics'' limited) or, more idealistically, is known 
perfectly up to statistical fluctuations (``statistics'' limited) (see Sec.~\ref{sec:Belle-II-projections} for details).  
The gray shaded region is excluded by combining LEP bounds~\cite{Fox:2011fx} with $g_\chi$-perturbativity.
For the hidden photon case, this limit is strengthened by including $Z$-pole 
constraints~\cite{Hook:2010tw} on $\varepsilon$, as shown by the green line.
See text for more details.}
\label{fig:heavy-off-shell}
\end{figure*}

%%%%%%%%%%%%%%%%%%%%%%%%%%%%%%%%%%%%%%%%%%%%%%%%%%%
\section{Constraints from \babar\ data}
\label{sec:BaBar}
%%%%%%%%%%%%%%%%%%%%%%%%%%%%%%%%%%%%%%%%%%%%%%%%%%%

The \babar\ Collaboration performed an unpublished 
analysis of mono-photon events in a search for decays of the $\Upsilon(3S)$ to $\gamma \, A^{0}$, where $A^{0}$ is an invisible pseudoscalar particle~\cite{Aubert:2008as}. 
 We reproduce their preliminary data in Fig.~\ref{fig:typicalsim}.  
 The search was performed on a sample of $122\times10^{6}$ $\Upsilon(3S)$ decays, corresponding to about $28$/fb of data at $\sqrt{s}\approx m_{\Upsilon(3S)}\approx10.355 \g$~\cite{Lees:2013rw}.
The data was analyzed in two overlapping photon CM energy regimes with distinct trigger requirements: $3.2 \g<E_\gamma^*<5.5 \g$ and $2.2 \g<E_\gamma^*<3.7 \g$, referred to respectively as the High-E and Low-E regions.  
The former used the full dataset, and the latter a subset corresponding to $19$/fb.
The main SM backgrounds are a peak at $m_{\chi \overline \chi}^2 = 0$ from $e^+ e^-\to \gamma \slashed \gamma$, a continuum background from $e^+ e^- \to \gamma \slashed e^+ \slashed e^-$, $e^+e^-\to \gamma \slashed \gamma \slashed \gamma$, where 
$\slashed e^\pm$ and $\slashed \gamma$ represent particles that escape undetected (down the beam pipe or in a detector crack) and, to a lesser extent, two-photon production of hadronic states decaying to photons where only one is detected. 
The results of a bump hunt in the photon spectrum were presented as 
 preliminary 
upper limits on the branching fraction (BF) ${\mathcal B}(\Upsilon(3S)\to\gamma A^0)\times{\mathcal B}(A^0\to inv.)$.

We use this data below to constrain the non-resonant production of LDM in $e^+ e^-$ 
collisions as shown in Fig.~\ref{fig:production}-{\em right}, for the three regions shown in Fig.~\ref{fig:search-regions}.
(A similar analysis is performed in Ref.~\cite{Yeghiyan:2009xc} to constrain LDM couplings to $b$-quarks through an effective dimension-6 operator.)
The \babar\ analysis applies both geometric and non-geometric cuts to the mono-photon data, with total efficiency for signal events given as 10-11\% (20\%) in the High-E (Low-E) region.
By simulating $e^+ e^-\to \Upsilon(3S) \to \gamma A^0$ events, we find that geometric acceptance accounts for 34\% and 37\% of this efficiency in the two respective regions, with non-geometric cuts therefore having about 30\% and 55\% efficiencies.
In our analysis, we determine the geometric cut acceptances for each search region from simulation, and apply a further cut 
of 30\% (55\%) in the High-E (Low-E) region to account for the efficiencies of other cuts.
Photon energies are smeared using a crystal-ball function, with tail parameters $\alpha=0.811$ and $n=1.79$, obtained from fitting the $E_\gamma^*$ distribution of $e^+ e^- \to \gamma \slashed \gamma$ to the data in~\cite{Aubert:2008as}. We take the width, $\sigma_{E_\gamma}/E_\gamma$, to be $1.5\%/(E_\gamma/\g)^{1/4} \oplus 1\%$ to match the values of $\sigma_{m_{\chi \overline \chi}^2}$ given in~\cite{Aubert:2008as}.
The signal was simulated with {\tt Madgraph 5}~\cite{Alwall:2011uj}.

%%%%%%%%%%%%%%%%%%%%%%%%%%%%%%%%%%%%%%%%%%%%%%%%%%%
\subsection{Constraints for Off-shell Heavy Mediators}
\label{sec:heavy-med}
%%%%%%%%%%%%%%%%%%%%%%%%%%%%%%%%%%%%%%%%%%%%%%%%%%%

When $\gamma + \chi \overline \chi$ events are produced through a heavy off-shell mediator (region (a) of Fig.~\ref{fig:search-regions}), the mono-photon spectrum has a shape very similar to that of the background, as can be seen in Fig.~\ref{fig:typicalsim}.
Because of this, and since no background 
estimate 
was performed in~\cite{Aubert:2008as}, we place constraints by requiring that the expected signal does not exceed the observed number of events by more than $2\sigma$ in any bin.

Fig.~\ref{fig:heavy-off-shell} (\textit{left}) shows lower limits on $m_{A'}/\sqrt{g_e g_\chi}$ as a function of $m_{A'}$ for a fixed DM mass of 10~MeV, while Fig.~\ref{fig:heavy-off-shell} (\textit{right}) shows limits as a function of $m_\chi$ for a fixed mediator mass of 12~GeV, and various mediators. The dependence on the type of mediator is negligible for $m_\chi \simlt 1$~GeV.  
The solid blue curves show projections for a similar search at Belle II (see Sec.~\ref{sec:Belle-II-projections}). These rely on the possibility of performing 
an estimate of the background 
and hence could also apply to a reanalysis of the data by the \babar\ collaboration if they are able to calculate the backgrounds and/or determine them from data.

These models are also constrained by mono-photon searches at LEP, which in this regime place an upper bound on $g_e$ (see Sec.~\ref{sec:LEP}). Combining this with the requirement $g_\chi<\sqrt{4\pi}$ (for perturbativity) gives the gray shaded region shown in Fig.~\ref{fig:heavy-off-shell}.
LEP's high CM energy makes it more effective at constraining heavier mediators, and the LEP bounds are stronger than those from \babar\ for $m_{A'}\simgt 15$~GeV. In the case of a hidden photon mediator, there is an even stronger constraint of $g_e < 0.026 e$ from $Z$-pole measurements~\cite{Hook:2010tw}, shown by the green lines.

\begin{figure}[t!]
\begin{center}
\includegraphics[width=0.48\textwidth]{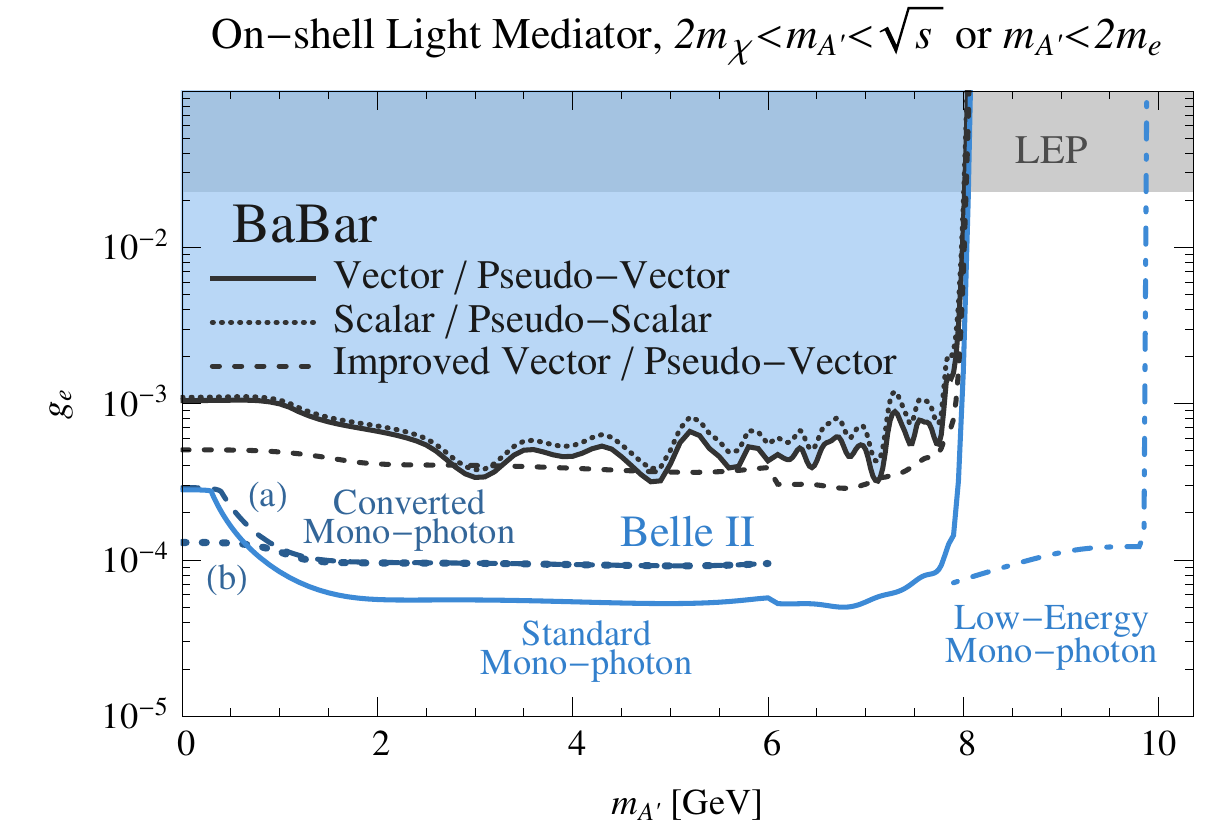}
\vspace{-5mm}
\caption{Upper bounds on the coupling of electrons to a mediator decaying invisibly to dark-sector states (region (b) of Fig.~\ref{fig:search-regions}). 
The solid black line / blue shaded region shows the bound from \babar\ data (this work), for a vector or pseudo-vector mediator. The dotted line shows the bound for a scalar or pseudo-scalar mediator. 
The  black dashed line shows the projected upper limit from an ``improved \babar" analysis for a vector or pseudo-vector mediator, where the $\gamma\slashed\gamma$ background has been reduced by a factor of 10. 
The projected reaches of four possible searches for a vector  
mediator at Belle II are shown by the solid blue lines: a converted mono-photon search (dashed, labelled (a) and (b), which respectively assume no (a factor of 10) improvement in the $\gamma\slashed\gamma$ background rejection over the ``improved \babar'' projection), 
a standard mono-photon search (solid), and a low-energy mono-photon search (dot-dashed) (see Sec.~\ref{sec:Belle-II-projections}).
The gray shaded region is excluded by LEP~\cite{Fox:2011fx}.  Additional limits relevant for sub-GeV mediators are shown in Fig.~\ref{fig:A'-invisible}.
See text for more details.
}
\label{fig:light-on-shell}
\end{center}
\end{figure}

%%%%%%%%%%%%%%%%%%%%%%%%%%%%%%%%%%%%%%%%%%%%%%%%%%%
\subsection{Constraints for On-Shell Light Mediators}
\label{sec:light-on-shell}
%%%%%%%%%%%%%%%%%%%%%%%%%%%%%%%%%%%%%%%%%%%%%%%%%%%

Production of on-shell invisible mediators in $e^+ e^- \to \gamma A'$ events (region ($b_{1,2}$) of Fig.~\ref{fig:search-regions}) gives a mono-photon signal with a distinct bump at $m_{\chi \overline \chi}^2 = m_{A'}^2$,
as illustrated in Fig.~\ref{fig:typicalsim}. 
The backgrounds are smooth functions, except for a bump at $m_{\chi \overline \chi}^2 = 0$ from $\gamma \slashed \gamma$ events.
We set limits on $g_e$ by performing our own bump hunt on the \babar\ data, as described below.

Following~\cite{Aubert:2008as}, we model the background in the High-E region by combining a crystal ball peaked at $m_{\chi \overline \chi}^2 = 0$ with an exponential $\exp(c \, m_{\chi \overline \chi}^2)$. In the Low-E region we combine an exponential $\exp(c_1 \, m_{\chi \overline \chi}^2 + c_2 \, m_{\chi \overline \chi}^4)$ with a constant. The normalizations of each component, and the exponents $c$, $c_1$, $c_2$, are treated as free nuisance parameters, with the normalizations constrained to be positive. 
We model the signal with a crystal ball peaked at $m_{\chi \overline \chi}^2 = m_{A'}^2$, and integrated area $N_{\rm signal}$.
The width of the crystal ball functions is as described above. 

For any given value of $m_{A'}$, we bin the expected rates using the same binning as the BaBar data, construct a likelihood function based on the signal and various background components, with the various nuisance parameters kept unconstrained 
except for the  
 normalizations, which are kept positive. We then set 95\% C.L. limits on $N_{\rm signal}$ using the profile likelihood method. 

The absence of features in the non-$\gamma \slashed \gamma$ backgrounds makes the bump-hunt an effective procedure to discriminate a signal from background for heavier $A'$. 
In the analysis of~\cite{Aubert:2008as}, 
only a limited background estimate was 
done 
on the $\gamma \slashed \gamma$ peak, using off-resonance data to estimate the background rate. 
We cannot use this approach in our analysis, since our signal would also appear in the off-resonance sample. The search becomes therefore background-limited for $m_{A'} \simlt 1$~GeV in the current \babar\ data. 
However, an improved background estimate may be possible. 
We therefore show a projection for an 
``improved \babar'' limit, 
assuming that the $\gamma \slashed \gamma$ background can be reduced by a factor of 10.  For this case, we fit smooth curves to the 
current \babar\ data to show the expected limit. 
 At Belle II, additional improvements in both background rejection and resolution may decrease the value of $m_{A'}$ 
at which the search becomes background-limited to a few hundred MeV, see Sec.~\ref{sec:Belle-II-projections}.  

We convert the limits on $N_{\rm signal}$ into limits on $g_e$ using simulation, accounting for the cut efficiency as described above. The limits are shown in Fig.~\ref{fig:light-on-shell}, along with projections for Belle II and limits from LEP (see Secs.~\ref{sec:Belle-II-projections} and~\ref{sec:LEP}).
In Figs.~\ref{fig:A'-invisible} and \ref{fig:A'-invisible-10-100-MeV} we show our limits in the $\varepsilon$ versus $m_{A'}$ plane for the special case of an invisibly decaying hidden photon. The bounds and projected reach of various other experiments are also shown, and are discussed further in Sec.~\ref{sec:hidden-photon-constraints}.

%%%%%%%%%%%%%%%%%%%%%%%%%%%%%%%%%%%%%%%%%%%%%%%%%%%
\subsection{Constraints for Off-Shell Light Mediators}
\label{sec:light-off-shell}
%%%%%%%%%%%%%%%%%%%%%%%%%%%%%%%%%%%%%%%%%%%%%%%%%%%

\begin{figure}[t!]
\begin{center}
\includegraphics[width=0.48\textwidth]{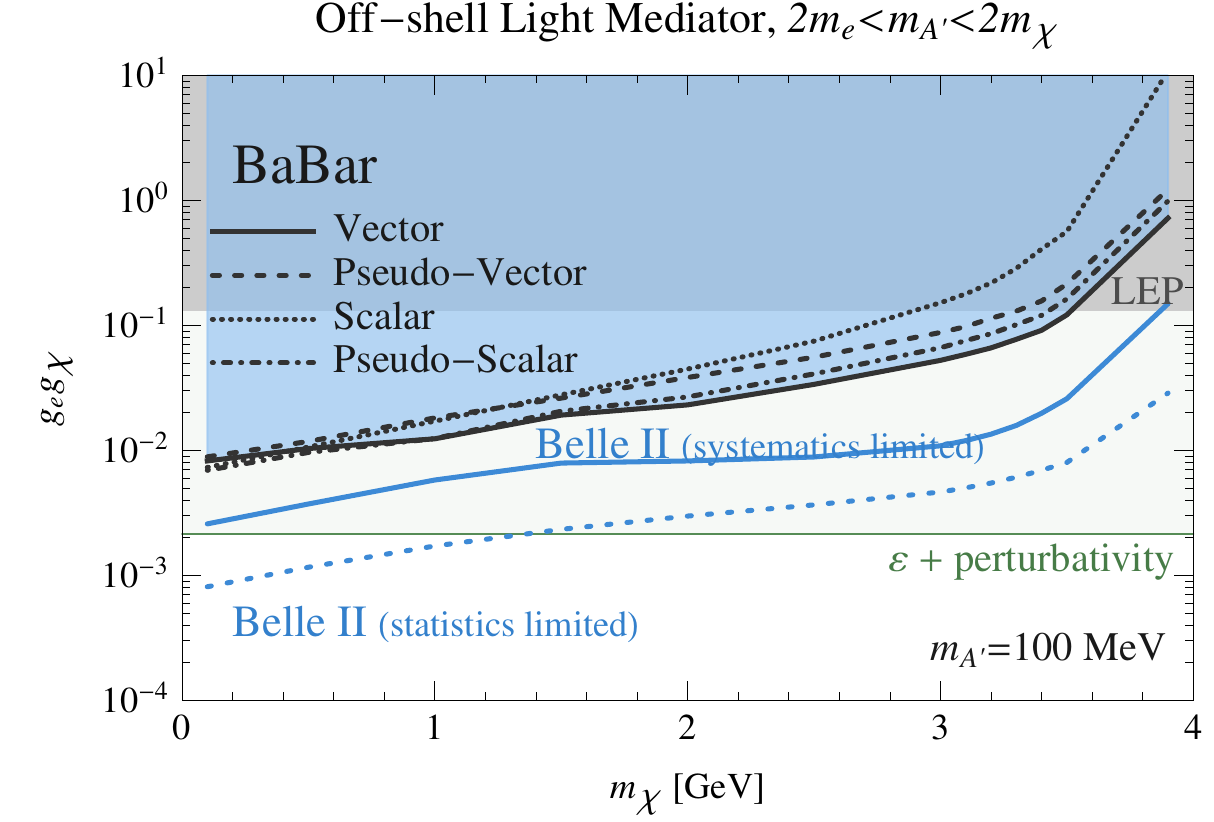}
\vspace{-5mm}
\caption{
Upper limits on $g_e  g_\chi$ for the  off-shell light mediator region (region (c) of Fig.~\ref{fig:search-regions}),
for a fixed mediator mass of 100~MeV.
The coloring and assumptions of the \babar\ and Belle II curves are as in Fig.~\ref{fig:heavy-off-shell}.
The gray shaded region is excluded by LEP~\cite{Fox:2011fx}.
With a hidden-photon mediator, there is a stronger constraint from combining $g_\chi$-perturbativity with a search for visibly-decaying hidden-photons at KLOE (green line). 
The possible reach of an edge search is not shown, but may allow some improvement.
The solid and dotted blue line both show the projected reach of Belle~II in the vector-mediated case assuming 
that the various background components are known at the $5-20\%$ level (``systematics'' limited) or, more idealistically, are known 
perfectly up to statistical fluctuations (``statistics'' limited) (see Sec.~\ref{sec:Belle-II-projections} for details). 
See text for more details.
}
\label{fig:light-off-shell}
\end{center}
\end{figure}

When $2m_e < m_{A'}< 2 m_\chi$ (region (c) of Fig.~\ref{fig:search-regions}), $\gamma + \chi \overline \chi$ production proceeds through a light off-shell mediator, giving a broad mono-photon spectrum as seen in Fig.~\ref{fig:typicalsim}. 
This spectrum has a kinematic edge at $m_{\chi \overline \chi}^2 = 4 m_\chi^2$.
Without good control over backgrounds, this spectrum is difficult to distinguish from backgrounds, and we conservatively place constraints by requiring that the expected signal does not exceed the observed number of events by more than $2\sigma$ in any bin.

Fig.~\ref{fig:light-off-shell} shows the upper limit on $g_e g_\chi$ as a function of $m_\chi$ for a fixed mediator mass of 100~MeV, for various mediator types. The constraint on $g_e g_\chi$ from LEP (see Sec.~\ref{sec:LEP}) is shown by the gray shaded region. In the case of a hidden photon mediator there is a stronger constraint, shown by the green line. This combines the requirement $g_\chi < \sqrt{4\pi}$ (for perturbativity) with bound on a visibly-decaying hidden photon by the KLOE experiment, which constrains $g_e < 0.002$ for $m_{A'}=100$~MeV~\cite{Archilli:2011zc}.
We note that if the mediator can decay to a second light state in the hidden sector then the visible constraints do not apply.  However, 
this second light state is then constrained by the on-shell constraints in Sec.~\ref{sec:light-on-shell}, which are of comparable strength. 

Also shown is the projected reach of Belle II for the vector-mediated case (see Sec.~\ref{sec:Belle-II-projections}). 
As for the heavy off-shell region, these rely on the possibility of performing 
 a background estimate
and hence could also apply to a reanalysis of the data by the \babar\ collaboration if control over the various background components can be obtained. In addition, a search for a kinematic edge may allow for an improvement of the bounds, but is not shown here.  
As can be seen from the figure, for the case of a hidden photon mediator, stronger constraints can be obtained from the direct production and (visible) decay of the $A'$.

%%%%%%%%%%%%%%%%%%%%%%%%%%%%%%%%%%%%%%%%%%%%%%%%%%%
\section{Comparison with other probes of Light Dark Matter}
\label{sec:comparison-to-LEP-DD}
%%%%%%%%%%%%%%%%%%%%%%%%%%%%%%%%%%%%%%%%%%%%%%%%%%%%%%

%%%%%%%%%%%%%%%%%%%%%%%%%%%%%%%%%%%%%%%%%%%%%%%%%%%
\subsection{LEP}
\label{sec:LEP}
%%%%%%%%%%%%%%%%%%%%%%%%%%%%%%%%%%%%%%%%%%%%%%%%%%%

The search for mono-photon events in $e^+ e^-$ collisions is also possible with the $\mathcal O(1)$/fb of data collected with a mono-photon trigger at LEP, and in~\cite{Fox:2011fx} this was used to place constraints on DM coupled to electrons through a mediator or a higher-dimension four-point interaction.
Because LEP operated at $\sqrt s \sim 200$~GeV, the DM and mediators of interest in this paper are light by LEP standards. There are therefore two regimes of interest to us: $2m_\chi < m_{A'} \ll 200$~GeV, for which the mono-photon signal rate is controlled by the single parameter $g_e$, and $m_{A'} < 2 m_\chi \ll 200$~GeV, for which it is controlled by the combination $g_e g_\chi$.

In Ref.~\cite{Fox:2011fx} (Fig.~7) bounds are presented in terms of the parameters $m_\chi$, $m_{A'}$, the effective cutoff  scale $\Lambda \equiv m_{A'}/\sqrt{g_e g_\chi}$, and the $A'$ decay width $\Gamma_{A'}$. However, in the two respective mass regimes of interest these are consistent with a single bound on either $g_e$ or $g_e g_\chi$, of
\begin{align}
g_e &< 0.023  \hspace{-40pt} &(2m_\chi < m_{A'}) \, , \\
g_e g_\chi &< 0.13  \hspace{-40pt} &(m_{A'} < 2 m_\chi) \, .
\end{align}
The former is extracted from the ``minimum width'' curves of Ref.~\cite{Fox:2011fx}, which correspond approximately to the assumption $g_\chi = g_e$.

In Figs.~\ref{fig:light-on-shell} and~\ref{fig:light-off-shell} we show these two bounds directly with  gray shaded regions. In Fig.~\ref{fig:heavy-off-shell} we show a combination of the bound on $g_e$ with the requirement $g_\chi < \sqrt{4\pi}$ (for perturbativity).
LEP is more suited to probing higher mass scales, and becomes more sensitive than \babar\ for mediator masses above about 15~GeV assuming $g_\chi=\sqrt{4\pi}$. 
For the  on- and off-shell  light mediator regimes, the bounds from \babar\ are significantly stronger than those from LEP, due largely to \babar's higher luminosity and larger production cross-section.

%%%%%%%%%%%%%%%%%%%%%%%%%%%%%%%%%%%%%%%%%%%%%%%%%%%
\subsection{Constraints on Hidden Photons $A'$}  
\label{sec:hidden-photon-constraints} 
%%%%%%%%%%%%%%%%%%%%%%%%%%%%%%%%%%%%%%%%%%%%%%%%%%%

In this subsection, we discuss various other probes related to the specific case of hidden photons that couple to light 
hidden-sector  states, possibly DM.  
Projections and/or constraints from these other probes are shown in 
Figs.~\ref{fig:A'-invisible} and \ref{fig:A'-invisible-10-100-MeV}, together with the $B$-factory constraints 
and projections shown already in Fig.~\ref{fig:light-on-shell}.  
We focus on rare kaon decays, precision measurements of the anomalous magnetic moment of the electron and muon, and fixed-target experiments.\footnote{We do not discuss or show a constraint from invisible $J/\psi$-decays~\cite{Ablikim:2007ek}, 
since it is much weaker than other constraints, except in a very narrow mass range near the $J/\psi$-mass --- see e.g.~\cite{Dharmapalan:2012xp}.}

%%%%%%%%%%%%%%%%%%%%%%%%%%%%%%%%%%%%%%%%%%%%%%%%%%%
\subsubsection{Rare Kaon decay limits involving an $A'$}
%%%%%%%%%%%%%%%%%%%%%%%%%%%%%%%%%%%%%%%%%%%%%%%%%%%

Meson decays involving hidden photons can constrain parts of the parameter space.  
A particularly important rare decay mode is $K^+ \to \pi^+ A'$, with $A' \to $ invisible.  
A search for the SM process $K^+ \to \pi^+ \nu \bar\nu$ by the BNL experiments E787~\cite{Adler:2004hp} 
and E949~\cite{Artamonov:2008qb}
found a total of seven events.  The SM predicted value~\cite{Brod:2010hi},
\bea
B_{\rm SM} (K^+ \to \pi^+ \nu \bar\nu) = (7.81 \pm 0.80) \times 10^{-11}\,,
\eea
is consistent with a combined result of E787 and E949 of the branching ratio measurement~\cite{Artamonov:2008qb}
\bea
B_{\rm measured} (K^+ \to \pi^+ \nu \bar\nu) = (17.3^{+11.5}_{-10.5}) \times 10^{-11}\,.
\eea

For the two-body decay $K^+ \to \pi^+ A'$ (where the $A'$ is on-shell), the $\pi^+$-momentum spectrum is peaked at 
\bea
|\vec{p}_{\pi}| 
 & = & \frac{1}{2 m_K} \, \big(m_K^4 + m_\pi^4 + m_{A'}^4 + \nonumber \\
 & & \hspace{0.5cm} - 2\, (m_K^2 m_\pi^2 + m_K^2 m_{A'}^2 + m_\pi^2 m_{A'}^2)\big)^{\f{1}{2}}\,,
\eea
while for three-body decay $K^+ \to \pi^+ A'^* \to \pi^+ \bar \chi \chi$ through an off-shell $A'$, the $\pi^+$-momentum has a continuous 
distribution, making it more difficult to distinguish from the SM decay $K^+ \to \pi^+ \nu \bar\nu$.  
The constraints are thus much stronger for the on-shell decay compared to the off-shell decay, and we only consider the former.

\begin{figure}[t!]
\begin{center}
\includegraphics[width=0.48 \textwidth]{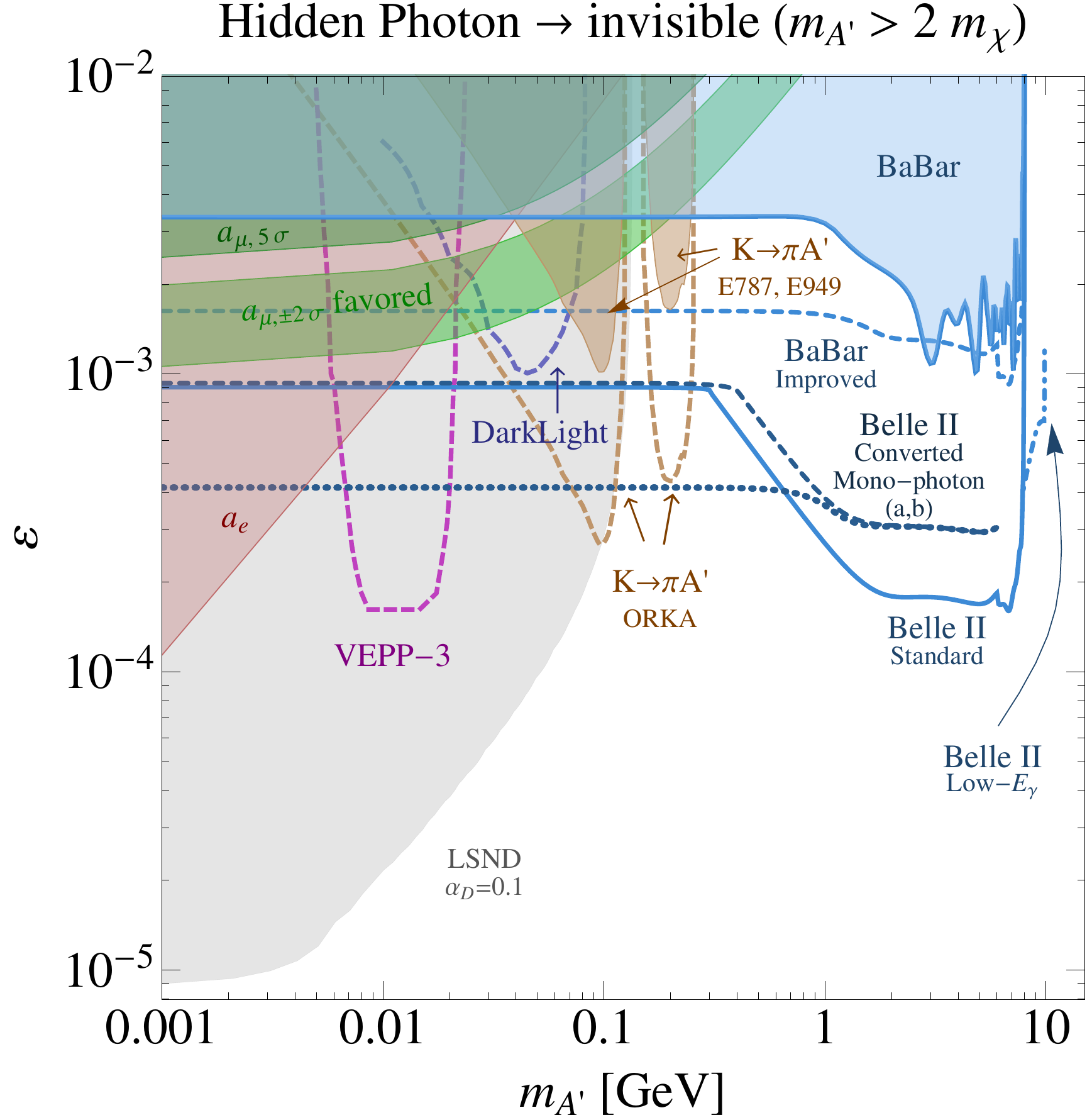}
\caption{
Constraints in the $\varepsilon$ versus $m_{A'}$ plane for {\bf invisibly-decaying} hidden photons.  
The bounds from the \babar\ mono-photon data are shown by the blue shaded region. 
The blue dashed line shows a ``\babar\ improved'' projection that assumes a factor of 10 reduction in the $\gamma\slashed\gamma$ 
background. 
Projections for four possible Belle II searches are shown by the four blue lines, with line styles matching 
Fig.~\ref{fig:light-on-shell} (see Sec.~\ref{sec:Belle-II-projections}): 
a converted mono-photon search (dashed, labelled (a) and (b), which respectively assume no (a factor of 10) improvement in the $\gamma\slashed\gamma$ background rejection over the ``\babar\ improved'' projection), 
a standard mono-photon search (solid), and a low-energy mono-photon search (dot-dashed) (see Sec.~\ref{sec:Belle-II-projections}).  
Various other constraints (shaded regions) and projected sensitivities (dashed lines) are also shown:
the anomalous magnetic moment of the electron ($a_e$, red) and muon ($a_\mu$, green),
rare kaon decays (brown), 
and the upcoming electron/positron fixed-target experiments DarkLight and VEPP-3. 
In the green shaded band an $A'$ could explain the discrepancy between the measured and predicted SM value of $a_\mu$. 
The gray shaded region is a constraint from LSND~\cite{deNiverville:2011it}, assuming $\alpha_D = 0.1$ and that  $\chi$ 
has no decay modes available to other light hidden-sector states that do not couple to the $A'$.   
More details are given in Sec.~\ref{sec:hidden-photon-constraints}, and we show the corresponding plot for $m_\chi = 10$~MeV and 100~MeV in Fig.~\ref{fig:A'-invisible-10-100-MeV} in Appendix~\ref{sec:invisible-A'}.
}
\label{fig:A'-invisible}
\end{center}
\end{figure}

Both E787 and E949 published results on the branching ratio limit for on-shell decays~\cite{Adler:2004hp,Artamonov:2009sz}, 
which we can use to constrain $\varepsilon$. 
Following~\cite{Pospelov:2008zw}, and using results from~\cite{DAmbrosio:1998yj}, the two-body decay width is given by 
\bea
&& \Gamma (K^+ \to \pi^+ A') = \f{\varepsilon^2 \alpha m_{A'}^2}{2^{10} \pi^4 m_K} \, \left|W(m_{A'}^2/m_K^2)\right|^2 \nonumber \\
& & \hspace{0.5cm} \times \left(1+ \f{(m_\pi^2 - m_{A'}^2)^2}{m_K^4} 
- \f{2 (m_\pi^2 + m_{A'}^2)}{m_K^2} \right)^{\f{3}{2}},
\eea
where 
\bea
\left|W(x)\right|^2 \simeq 10^{-12}\,\left(3 + 6 x \right)\,.
\eea
Using now the measured total width of the $K^+$ of $\Gamma_{\rm total} (K^+) \simeq 5.3\times10^{-14}$~MeV, and  
taking the E949 limit on the branching ratio $K^+\to \pi^+ A'$ from Fig.~18 in~\cite{Artamonov:2009sz} (scaled to 95\% C.L.), 
we derive the limit on $\varepsilon$ versus $m_{A'}$ shown in the shaded brown 
region in Figs.~\ref{fig:A'-invisible}  and \ref{fig:A'-invisible-10-100-MeV}.  
 There are two separated excluded regions (as opposed to a single continuous region), since the search 
$K^+\to \pi^+\bar \nu\nu$ was restricted to certain values of $|\vec{p}_\pi|$ to avoid backgrounds.

Several experiments have been proposed with an improved sensitivity to $K^+ \to \pi^+ \bar\nu\nu$ decays.  
ORKA~\cite{E.T.WorcesterfortheORKA:2013cya} is a proposed experiment to measure this branching ratio to much 
higher precision using stopped kaons from the Fermilab Main Injector high-intensity proton source.  
Its detector design is based on the E787 and E949 experiments, and it is expected to detect $\sim 1000$ decays over five years of 
data taking, improving the branching ratio measurement to 5\%.  ORKA is expected to be able to take data five years after funding 
becomes available.  
A rough sensitivity estimate of ORKA to $K^+\to \pi^+ A'$ decays can be obtained by scaling the E949 limit 
in~\cite{Artamonov:2009sz} used above.  
First, we assume a factor of 100 increase in the luminosity.  In addition, we assume that the background rate of $K^+ \to \pi^+ \bar\nu\nu$ 
decays agrees with the SM prediction (in E787 and E949 the observed background rate was found to be twice as large as the SM 
prediction, but still consistent with it, thereby weakening the limits slightly).  ORKA can thus be expected to improve the branching ratio limit 
by at least $\sim \sqrt{200}\sim 14$, and improve the sensitivity to $\varepsilon$ by $\sqrt[4]{200}\sim 3.8$, which is shown 
in Figs.~\ref{fig:A'-invisible} and \ref{fig:A'-invisible-10-100-MeV} with dashed  brown 
 lines.  
Note that this ignores expected improvements in the $\pi^+$-momentum resolution.  
This projected improvement in sensitivity to the branching ratio is also weaker than what is projected by the ORKA collaboration 
for $m_{A'}=0$, namely from $0.73\times 10^{-10}$ (at 90\% C.L.) to $2\times 10^{-12}$, a factor of 36.5 as opposed to 14 
(see e.g.~\cite{ORKA-talk}).  The ORKA sensitivity shown in Figs.~\ref{fig:A'-invisible} and \ref{fig:A'-invisible-10-100-MeV} should 
thus be viewed as conservative.  

Another experiment with excellent sensitivity to $K^+ \to \pi^+ \bar\nu\nu$ decays is NA62 at CERN 
(with $\sim 50$ events/year)~\cite{NA62}.  
NA62 may begin data taking within a year.  It uses decay-in-flight kaons and may be sensitive to lower $\pi^+$-momenta and 
thus slightly higher $m_{A'}$.  We do not show a sensitivity estimate for NA62, although it would 
be interesting for the NA62 collaboration to look at this decay mode in detail.  
Finally, we note that a future Project X facility could reach even higher sensitivity than ORKA or NA62~\cite{Hewett:2012ns}.

%%%%%%%%%%%%%%%%%%%%%%%%%%%%%%%%%%%%%%%%%%%%%%%%%%%
\subsubsection{QED Precision Measurements}
%%%%%%%%%%%%%%%%%%%%%%%%%%%%%%%%%%%%%%%%%%%%%%%%%%%

An $A'$ contributes to the anomalous magnetic moment of the muon and electron, $a_e\equiv(g-2)_e$ and 
$a_\mu\equiv(g-2)_\mu$~\cite{Pospelov:2008zw}.  
For $a_\mu$, this contribution could resolve the long-standing disagreement between the calculated (see e.g.~\cite{Davier:2010nc}) 
and experimentally measured value~\cite{Bennett:2006fi}.  
The SM and measured values are 
\bea
a_{\mu}^{\rm SM} & = & (11659180.2 \pm 4.2 \pm 2.6 \pm 0.2) \times 10^{-10}\,, \\
a_\mu^{\rm exp} & = & (11659208.9 \pm 5.4 \pm 3.3) \times 10^{-10}\,,
\eea
and hence the difference 
\bea
\Delta a_{\mu} = a_\mu^{\rm exp} - a_\mu^{\rm SM} = (28.7 \pm 8.0) \times 10^{-10}
\eea
is about $3.6 \sigma$.  
In Figs.~\ref{fig:A'-invisible} and \ref{fig:A'-invisible-10-100-MeV}, we show the ``$2 \sigma$'' region in which an $A'$ helps solve this disagreement by contributing 
$a_{\mu}^{A'} = (28.7 \pm 16.0) \times 10^{-10}$.  We also show a ``$5\sigma$'' line, where the $A'$ contributes ``too much'',  
$a_{\mu}^{A'} = 68.7 \times 10^{-10}$.  

Current measurements of $a_e$ agree well with SM theory ~\cite{Aoyama:2012wj,2011PhRvL.106h0801B}  
and experiment~\cite{2008PhRvL.100l0801H}.   One finds (see also~\cite{Davoudiasl:2012ig,Endo:2012hp}),  
\bea\label{eq:delta-ae}
\Delta a_e = a_e^{\rm exp} - a_e^{\rm SM} = (-1.06\pm 0.82) \times 10^{-12}\,.
\eea 
The contribution from an $A'$ would introduce a   disagreement between the theory and experimental value. 
In Figs.~\ref{fig:A'-invisible}  and \ref{fig:A'-invisible-10-100-MeV}, we show the shaded region labelled $a_e$ in which 
$a_e^{A'} > (-1.06+ 3 \times 0.82) \times 10^{-12} = 1.4 \times 10^{-12}$. 

%%%%%%%%%%%%%%%%%%%%%%%%%%%%%%%%%%%%%%%%%%%%%%%%%%%
\subsubsection{Fixed-target and Beam-dump experiments}
%%%%%%%%%%%%%%%%%%%%%%%%%%%%%%%%%%%%%%%%%%%%%%%%%%%

Several existing and proposed experiments are sensitive to visible $A'$ decays, usually to 
$e^+e^-$ (see e.g.,~\cite{Bjorken:2009mm,Reece:2009un,Batell:2009yf,Essig:2009nc,Freytsis:2009bh,Batell:2009di,:2009pw,Bossi:2009uw,Essig:2010xa,Essig:2010gu,HPS,Archilli:2011zc,Abrahamyan:2011gv,Merkel:2011ze,Archilli:2011nh}).   These searches were motivated in part by astrophysical anomalies connected to Weak-scale DM~\cite{ArkaniHamed:2008qn,Pospelov:2008jd}.  
However, if the $A'$ can decay to light hidden-sector states, then many of these experiments lose all their sensitivity.  There are some exceptions, including the electron/positron fixed-target experiments DarkLight~\cite{Freytsis:2009bh,DarkLight} and VEPP-3~\cite{Wojtsekhowski:2012zq}, which have sensitivity also to invisible $A'$ decays by performing a missing mass measurement.  
In Fig.~\ref{fig:A'-invisible}, for DarkLight, we show the reach as shown in Fig.~18 of~\cite{Kahn:2012br} (for 95\% photon efficiency); 
for VEPP-3, we show the reach given in~\cite{Wojtsekhowski:2012zq}.

Other experiments sensitive to invisible $A'$ decays include proton fixed-target experiments, in which a proton beam incident on a target 
produces a large number of mesons that decay to an $A'$ (e.g.~$\pi^0 \to \gamma A'$), which in turn decays to LDM~\cite{Batell:2009di}.  LSND in particular provides strong, but model-dependent, constraints~\cite{deNiverville:2011it}, which we show with gray shaded 
regions on Figs.~\ref{fig:A'-invisible} and \ref{fig:A'-invisible-10-100-MeV}.
Further searches are possible at several neutrino facilities~\cite{Batell:2009di,deNiverville:2011it,deNiverville:2012ij}, and 
a proposal has been submitted to the MiniBooNE Collaboration~\cite{Dharmapalan:2012xp}.   
We do not show the reach for these experiments on our plots.
Note that the constraints on $\varepsilon$ are proportional to $\sqrt[4]{\alpha_D}$, and thus disappear for small $\alpha_D = g^2_\chi/4\pi$.  
The constraints from these experiments also disappear if $\chi$ can decay into lighter hidden-sector particles that do not interact 
with the $A'$.

\subsubsection{Supernova}

An $A'$ can increase the cooling rate of supernovae.   Visible decays, with e.g.~$A'\to e^+e^-$, $\varepsilon\sim 10^{-10}-10^{-7}$, 
are constrained for $m_{A'}\lesssim 100-200$~MeV due to the cooling constraints on SN1987A~\cite{Bjorken:2009mm,Dent:2012mx} (see also \cite{Davidson:2000hf}).  
There is no bound for very small $\varepsilon$, since not enough $A'$ are produced to contribute significantly to the cooling. For  
larger $\varepsilon$, the $A'$ lifetime becomes short enough for it to decay inside the supernova, and so does not contribute to any 
cooling.  

For the case where $A'$ predominantly decays to LDM or other hidden-sector states, the situation is more complicated.    
As in the previous case, there is no constraint if $\varepsilon$ is small enough, since not enough $A'$s are produced.  
For larger $\varepsilon$, $A'$ decay to LDM inside the supernova, and there is potentially a bound if the LDM can escape the supernova.   
While a careful calculation of the supernova bound is beyond the scope of this paper, 
we make a few remarks below to evaluate their relevance to the region probed by \babar\ and Belle II.  
A dedicated discussion of the  bound will appear in~\cite{Dreiner:SN}.  
(We note that there are also constraints from white-dwarf cooling, but only if the $A'$ decays to LDM states with a 
mass $\lesssim 1$~keV~\cite{Dreiner:2013tja}.)

The mean free path of the LDM is given by $\ell \sim 1/ n \sigma$, where 
$n\simeq  2\times 10^{38}$/cm$^3$ is the number density of nucleons or electrons in a supernova.  
The cross section for LDM to scatter off an electron is 
\begin{equation}
\sigma_{e\chi\to e\chi} \simeq \left\{ \begin{array}{lll}
{\alpha g_\chi^2\varepsilon^2}/{m_{A'}^2} && m_{A'}\ll T
\\
{\alpha g_\chi^2 \varepsilon^2 T^2}/{m_{A'}^4} && m_{A'}\gg T
\end{array}\right.  \,.
\end{equation}
For a typical supernova temperature of $T\simeq 30$ MeV, $g_\chi \varepsilon = 10^{-4}$ and $m_{A'}=10$ MeV, the free-streaming length 
is $R_{\rm FS} \simeq 10^{-11}$ km, while for significantly larger masses, $m_{A'}\gtrsim 10$ GeV, one finds $R_{\rm FS} \simeq 10$~km. 
For most of the parameter space that can be probed by \babar\ and Belle II, the free-streaming length is thus 
smaller than the supernova radius. 

The production rate of LDM through an $A'$ is proportional to $\varepsilon^2$.  
In addition, for a large parameter region of relevance to \babar\ and Belle II, $m_{A'}\gg T$, so that the production is through an off-shell 
$A'$ and thus also proportional to $1/m_{A'}^4$.  
Ignoring the diffusion of LDM through the supernova, the cooling occurs via the escaping LDM produced close to the free-streaming surface.   For small free-streaming lengths, the overall energy that escapes (proportional to the production rate times the free-streaming length) is therefore independent of $\varepsilon$ and $m_{A'}$.  However, both the production rate and free-streaming length depend strongly on the density and temperature profiles, which are highly uncertain very close to the edge of the supernova.  Thus the computation of the cooling rate suffers from very large uncertainties and for small free-streaming lengths cannot be used to place a 
 robust limit.

When the free streaming length becomes of order the  size of the supernova, i.e.~for $m_{A'}\gtrsim 10$ GeV and  sufficiently small $\varepsilon$, any DM particles that are produced in the supernova will escape.  
However,  the temperature profile and  size of a supernova, and of SN1987A in particular (which is the only available data), is not known precisely, so that the 
precise value of $m_{A'}$ at which the free-streaming length equals the supernova radius is not known.  
For instance, for the same $g_\chi\epsilon=10^{-4}$ as above and under the assumption of a homogeneous temperature, $T_{\rm SN}=25$ MeV,  throughout the supernova, one finds $R_{\rm FS}= R_{\rm SN} = 10$ km for $m_{\rm A'} = 8.5$ GeV.  For $g_\chi\epsilon = 5\times 10^{-4}$, $T_{\rm SN} = 40$ MeV and $R_{\rm FS} = R_{\rm  SN} = 20$ km, the required $A'$ mass is $m_{A'} = 35$ GeV. 
Thus it is not clear if the supernova bounds apply at all to the $A'$ masses and couplings that can be probed by \babar\ or Belle II.

\begin{figure*}[t!]
\centering

\includegraphics[width=0.48\textwidth]{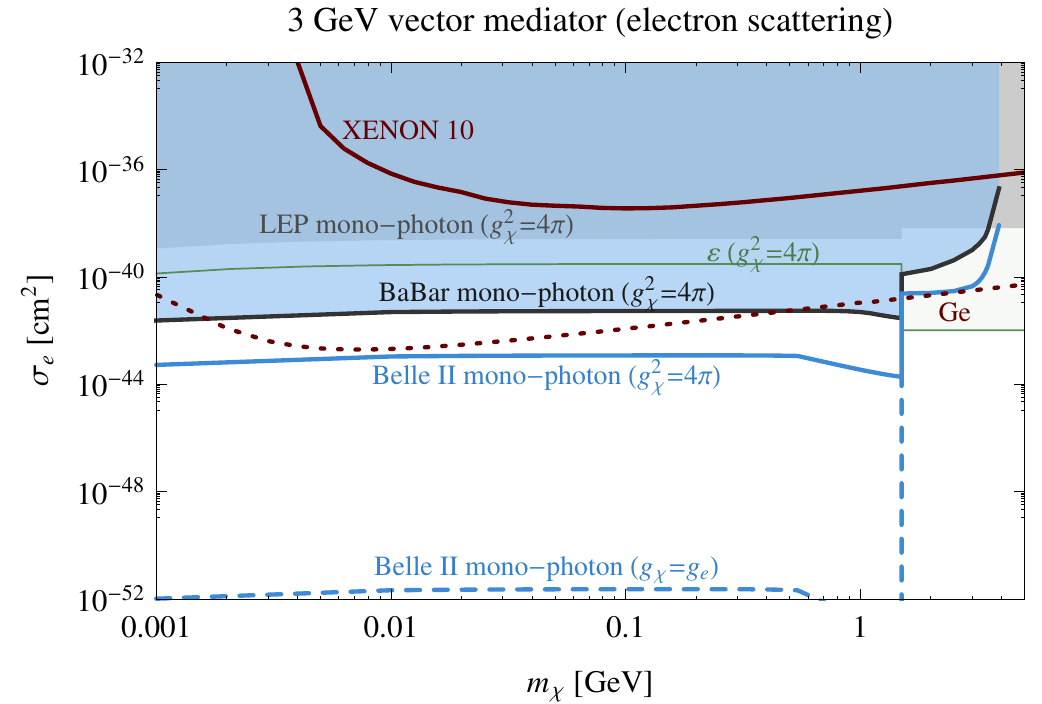}
\hfill
\includegraphics[width=0.48\textwidth]{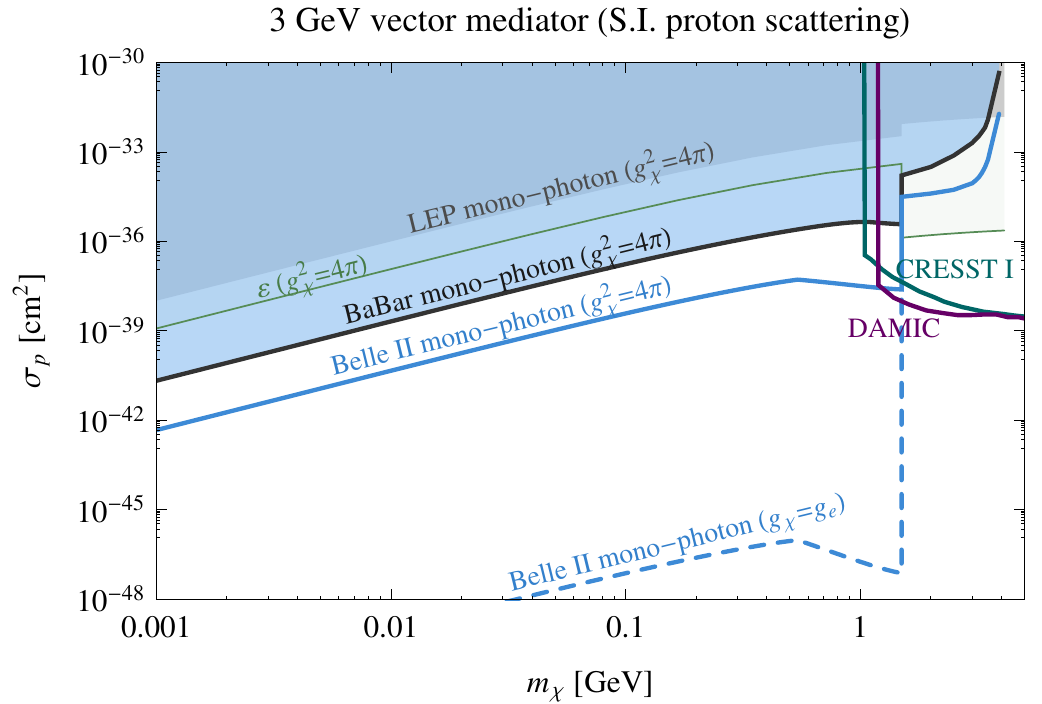}

\caption{A comparison of the sensitivities of mono-photon searches and direct detection experiments to LDM, taking a 3~GeV vector mediator for illustration. Constraints and projections are shown on the non-relativistic scattering cross-section of DM with electrons (\textit{{\bf left}}), and with protons (\textit{{\bf right}}), assuming for the latter that the coupling of the mediator to SM particles is proportional to their charge (as with a hidden photon).
Existing direct detection bounds on proton scattering: CRESST~\cite{Altmann:2001ax} (solid turquoise) and DAMIC~\cite{Barreto:2011zu} (solid purple), are shown. For electron scattering we show the XENON10 limit~\cite{Essig:2012yx} (solid dark red).  In addition, the dotted dark red line shows a projection for a germanium-based electron recoil experiment~\cite{Essig:2011nj}.
The constraint from \babar\ (this work) is shown as solid black line / blue shaded region. 
The discontinuity at $m_{\chi} = 1.5$~GeV corresponds to the transition between on-shell to off-shell light mediator regimes. In the latter regime we fix  $g_\chi = \sqrt{4\pi}$ (smaller $g_\chi$ would correspond to stronger bounds). 
LEP mono-photon searches~\cite{Fox:2011fx} are shaded in gray and limits from precision hidden photon searches are shown by the thin green line labeled ``$\varepsilon$'' (see text). 
For the projected reach of Belle II mono-photon searches (blue lines) we use the ``systematics limited'' bound for $m_\chi > 1.5$~GeV, and otherwise the stronger of the ``converted'' and ``standard'' mono-photon searches shown in Fig.~\ref{fig:A'-invisible} (see Sec.~\ref{sec:Belle-II-projections}). For $m_\chi < 1.5$~GeV we also show the projected reach of Belle II assuming $g_\chi = g_e$ (the boundary between visibly- and invisibly-decaying mediators).}
\label{fig:direct-detection-1}
\end{figure*}

%%%%%%%%%%%%%%%%%%%%%%%%%%%%%%%%%%%%%%%%%%%%%%%%%%%
\subsection{Direct Detection}
\label{sec:direct-detection}
%%%%%%%%%%%%%%%%%%%%%%%%%%%%%%%%%%%%%%%%%%%%%%%%%%%

Elastic nuclear recoils from DM scattering in current direct detection experiments are not able to probe DM with masses 
below a few GeV.  
However, it has recently been demonstrated that direct detection experiments can probe LDM below the GeV scale if the 
DM scatters instead off electrons~\cite{Essig:2011nj,Essig:2012yx}.   
While the limits in~\cite{Essig:2012yx} were derived from only a small amount of data taken with an experiment focused on probing heavier DM, near-future experiments such as CDMS, LUX, DAMIC and XENON100 are expected to significantly improve their sensitivity in upcoming years.   
Such constraints are somewhat complementary to the ones derived here, due to their sensitivity to distinct kinematical regimes.   Nonetheless, under certain assumptions, the constraints can be directly compared.  

In order to make such a comparison, we assume below that the mediator mass is larger than the typical momentum transfer, $q \sim $ keV,  relevant for electron scattering in direct detection experiments.  Consequently,  
effective operators of the form of Eqs.~\eqref{eq:eff1}-\eqref{eq:eff4} can be used to describe the relevant interactions at direct detection experiments, where the cut-off scale, $\Lambda$, is once again taken to be  $\Lambda = m_{A'}/\sqrt{g_e g_\chi}$.
For a sufficiently heavy mediator, corresponding to region (a) of Fig.~\ref{fig:search-regions}, the limit on $\Lambda$ is directly obtained from Fig.~\ref{fig:heavy-off-shell}. 
For lighter mediators, the mass $m_{A'}$ must also be specified.
For a mediator in region (c), this is combined with the \babar\ limit on $g_e g_\chi$ from Fig.~\ref{fig:light-off-shell} to set a bound on $\Lambda$.
For a light invisible mediator, in region ($b_{1,2}$), the bound on $g_e$ from Fig.~\ref{fig:light-on-shell} applies for any value of $g_\chi$ 
in the range $g_e \simlt g_\chi \simlt \sqrt{4 \pi}$.
To set a bound on $\Lambda$, we conservatively fix $g_\chi = \sqrt{4 \pi}$, at the limit of perturbativity.

Under the above assumptions, the DM-electron cross-section is simply given by 
\begin{equation}
\sigma_e=Q\frac{\mu_{\chi e}^2}{\pi\Lambda^4}\,.
\label{eq:dde}
\end{equation}
Here $Q=1$ for the vector and scalar mediator while $Q=3$ for the pseudo-vector mediator. $\mu_{\chi e}$ stands for the DM--electron reduced mass.
In the case of a pseudo-scalar mediator, direct detection rates are velocity suppressed and hence are not shown.
We also do not show the results for a pseudo-vector and scalar; for a scalar, there is no generic expectation for the mediator couplings 
when they are not proportional to the fermion yukawa couplings.  

 The \babar\ results may also be translated to DM-proton scattering rates, under assumptions about how the mediator couples to quarks.
For a vector mediator, motivated by kinetic mixing with the photon, we assume that the couplings are proportional to the electric charge of the SM particles.
One then finds the  cross-section for a vector mediated DM-proton interaction to be,
\begin{equation}
\sigma_p=\frac{\mu_{\chi p}^2}{\pi\Lambda^4}\,,
\end{equation}
where $\mu_{\chi p}$ stands for the reduced DM-proton mass.

\begin{figure*}[t!]
\centering

\includegraphics[width=0.48\textwidth]{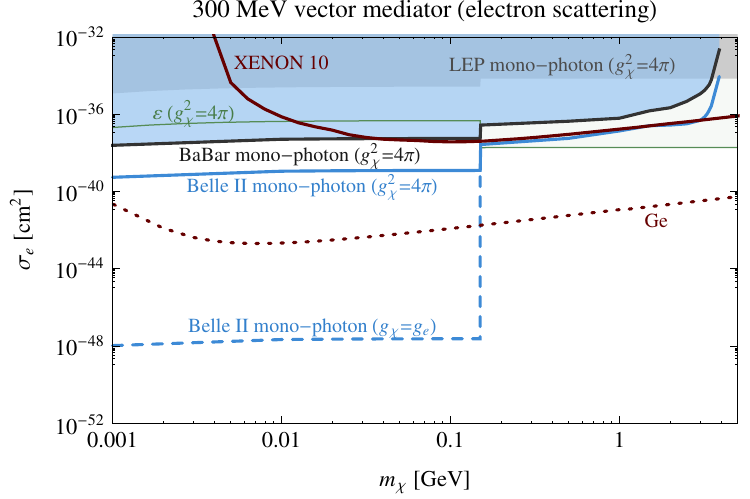}
\hfill
\includegraphics[width=0.48\textwidth]{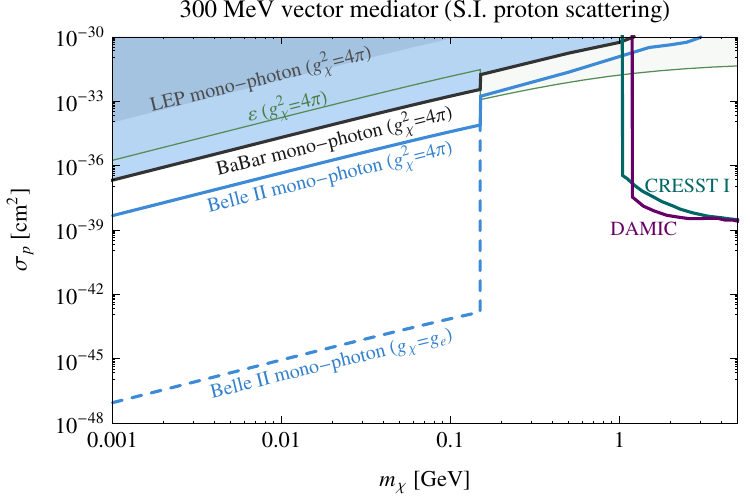}

\caption{A comparison of mono-photon searches and direct detection experiments, as in Fig.~\ref{fig:direct-detection-1}, but for a mediator mass of 300 MeV.}
\label{fig:direct-detection-2}
\end{figure*}

The resulting  direct detection limits on  the non-relativistic scattering cross-section of DM with electrons and with protons from the \babar\ search are shown in Figs.~\ref{fig:direct-detection-1} and~\ref{fig:direct-detection-2} for the case of a vector mediator. 
Results are shown as a function of $m_{\chi}$, for fixed mediator masses of 3~GeV and 300~MeV.
The plots are discontinuous for $ m_{\chi}= m_{A'}/2$, due to the transition between an on-shell and off-shell mediator. In addition, constraints from existing experiments are presented.  The LEP mono-photon searches~\cite{Fox:2011fx} are shown for comparison as a gray shaded region. The green line shows other bounds on a hidden photon mediator: 
$\varepsilon < 0.026$ for $m_{A'} = 3$~GeV from precision $Z$-pole measurements~\cite{Hook:2010tw};
$\varepsilon < 0.01$ for $m_{A'} = 300$~MeV from muon $(g-2)$ constraints (see Sec.~\ref{sec:hidden-photon-constraints}); 
and for $m_\chi > m_{A'}/2$, $\varepsilon \simlt 1.5 \times 10^{-3}$ from a search for $\Upsilon \to \gamma \mu^+ \mu^-$ by \babar\ \cite{Bjorken:2009mm}.
Finally, limits from DM direct detections experiments are also shown: the XENON10 limits~\cite{Essig:2012yx} (solid dark red) are presented for DM-electron scattering, while limits from CRESST~\cite{Altmann:2001ax} (solid turquoise) and DAMIC~\cite{Barreto:2011zu} (solid purple) are shown in the DM-proton case. The dotted dark red line shows a possible projection for a germanium-based electron recoil experiment~\cite{Essig:2011nj}, assuming a 1~kg-year exposure and no backgrounds. 

In each of the plots we demonstrate the projected Belle-II sensitivity as discussed in Sec~\ref{sec:Belle-II-projections}.  For $m_\chi > m_{A'}/2$ we show the ``systematics limited'' bound for a light off-shell mediator, while for $m_\chi <  m_{A'}/2$ we show the stronger of the projections for a converted mono-photon search and a standard mono-photon search shown in Figs.~\ref{fig:light-on-shell} and~\ref{fig:A'-invisible}.
%the reach for a converted mono-photon search with a 10$\times$ reduced $\gamma \slashed \gamma$ peak compared to the current \babar\ data. 
In the latter case, the solid blue line shows the conservative assumption $g_\chi = \sqrt{4\pi}$. The dashed blue line assumes $g_\chi = g_e$ (the boundary between visibly- and invisibly-decaying mediators), illustrating how much more powerful mono-photon searches are than direct detection experiments at constraining the hidden photon with a small $g_\chi$ scenario. 

The results above demonstrate the strength of low-energy collider experiments in searching for DM in regimes where direct detection experiments are still lacking.  Results from the future Belle-II experiment and from future direct detection searches (along the lines suggested in~\cite{Essig:2011nj,Essig:2012yx}) are competitive (although complimentary) for the case of a heavy mediator.  For a light mediator, direct detection experiments are expected to be crucial as their sensitivity is significantly better due to the distinct kinematics.

%%%%%%%%%%%%%%%%%%%%%%%%%%%%%%%%%%%%%%%%%%%%%%%%%%%%%%
\section{Projections for Belle II}
\label{sec:Belle-II-projections}
%%%%%%%%%%%%%%%%%%%%%%%%%%%%%%%%%%%%%%%%%%%%%%%%%%%%%%

The bounds placed on LDM by the \babar\ data are competitive with existing constraints, and in many cases stronger. 
In particular, for on-shell invisible mediators the bounds exclude a large region of previously-allowed parameter space.
In this section, we make projections for the sensitivity to LDM of future high-luminosity $e^+ e^-$ colliders, notably the 
next-generation $B$-factory Belle II, which could significantly improve on these results. 

Belle II is an upgrade of the Belle experiment, using the SuperKEKB asymmetric $e^+ e^-$ collider currently 
under construction~\cite{Abe:2010sj}. It is expected to start taking data in 2016 and obtain 50/ab of integrated luminosity at $\sqrt s \approx10.5$~GeV by 2022~\cite{Belle-II-talk:2013}.
The average energy resolution is slightly improved over \babar\, $\sigma_{E_\gamma}/E_\gamma=1.7\%$. 
The implementation of a mono-photon trigger will require a dedicated study by the Belle II collaboration to ensure that 
the high luminosity and pile-up do not lead to an unacceptably high trigger rate.  
It will hopefully be possible to implement such a trigger for the full Belle II run for energetic mono-photons without prescaling, 
and possibly with a prescaled version also at lower photon energies ($E_\gamma \lesssim 2~$GeV). 
An interesting possibility is the study of a dedicated trigger for mono-photons that convert in the tracker. 
While paying a high price in signal rate, this may overcome significant issues (discussed below) with the standard mono-photon search.

A precise estimate of the reach of Belle II depends on various unknowns, such as the amount of data taken with the mono-photon trigger, the trigger and cut efficiencies, and most crucially the background rate and the ability to perform a careful background estimate. 
In order to make illustrative projections, we make plausible assumptions about these factors.
We emphasize that the actual reach of Belle II may prove somewhat different than our projections.

\subsection{Standard Mono-Photon Search}
\label{sec:Belle-II-standard-projection}

We assume that Belle II can implement a mono-photon trigger on the full 50/ab of data, but restricted to the energy range $2.2\g < E_\gamma^* \simlt 5.5\g$, corresponding to $-5 \g^2 < m_{\chi \overline \chi}^2 \simlt 62\g^2$. This avoids the excessive rates coming from the radiative Bhabha  contributions and other soft photon and instrumental backgrounds at lower energies.
In analogy with the current \babar\ search~\cite{Aubert:2008as}, we divide the energy range again into High-E and Low-E regions. This allows us to scale up the current \babar\ background estimates. 
We determine these by fitting the \babar\ data with the smooth functions described in~\ref{sec:light-on-shell}.
We then assume the same geometric acceptance in each region as \babar , neglecting ${\cal O}(1)$ differences between the geometric acceptances of the two experiments due to differences in the beam energies and calorimeter rapidity coverage. We further assume a constant efficiency for non-geometric cuts of 50\% 
and finally scale up 
all the smoothed \babar\ backgrounds by the ratio of the luminosities, except for the $\gamma\slashed\gamma$ background.  
For the latter, we assume a baseline improvement in the background rejection by a factor of 10 over the current \babar\ analysis~\cite{Aubert:2008as} (so that it corresponds to the ``improved \babar'' version discussed in Sec.~\ref{sec:light-on-shell} and shown 
in Figs.~\ref{fig:light-on-shell} and \ref{fig:A'-invisible}) before scaling by the ratio of luminosities. 
We note that the real backgrounds may of course differ significantly from these simple scalings using the \babar\ data.  

 For on-shell mediators, we set limits following the same procedure described in Sect.~\ref{sec:BaBar}.
For off-shell mediators, we assume the expected continuum background rate can be determined using some combination of Monte Carlo and data-driven techniques, allowing one to estimate and effectively ``subtract'' part of the background 
(there would be no improvement in the Belle II sensitivity over the current \babar\ limits without an improved understanding of the backgrounds, since we showed ``signal-only'' limits for \babar).  
Without a realistic estimate of the range of shape variations for the various background components, it is hard to estimate the power of the limits that can be obtained by such subtraction procedure. Instead, we provide a conservative and an aggressive estimate of the limits, labeled ``systematics-limited'' and ``statistics-limited'', respectively. In both cases, we present single-bin limits, where the signal is constrained not to exceed the 95\% C.L. in any single bin. In the ``systematics-limited'' case, we assume that the bin uncertainties are dominated by systematic uncertainties due to the subtraction procedure: we set them at 5\% for the radiative Bhabha, 10\% for the $\gamma\slashed\gamma$ peak (whose size assumes that Belle II can reach the same level of rejection as the \babar\ analysis) and at 20\% for the other continuum components. In the ``statistics-limited'' case, we consider only statistical uncertainties. While it is almost impossible 
to achieve this limit, good control of the background shapes and the simultaneous fit of many bins could yield a significant improvement over our ``systematics-limited'' reach, so that it is instructive to show both.

The sensitivity reach from both these estimates are shown by the solid and dashed blue curves in 
Figs.~\ref{fig:heavy-off-shell} and \ref{fig:light-off-shell}, while only the ``systematics-limited'' reach is shown with the 
solid blue curves in Figs.~\ref{fig:light-on-shell}, \ref{fig:A'-invisible}, and \ref{fig:A'-invisible-10-100-MeV}.  
In Figs.~\ref{fig:direct-detection-1} and~\ref{fig:direct-detection-2}, the ``systematics-limited'' curve is shown under two assumptions for the value of $g_\chi$.

\subsection{Low-Energy Mono-Photon Search}
\label{sec:Belle-II-low-energy-projection}

We also consider the possibility that Belle II can implement a prescaled trigger for low-energy mono-photons, 
$0.5 \g < E_\gamma^* \le 2.2\g$.  We will assume a prescale factor of 10 (corresponding to 5/ab of collected data), 
although a dedicated study by the Belle II collaboration 
is necessary to see whether this is sufficient to avoid background events overwhelming the data acquisition.  
We estimate the background by extrapolating the fit in the Low-E region described in the previous section, Sec.~\ref{sec:Belle-II-standard-projection}, and otherwise follow the same assumptions and procedures.  The result is shown by the blue dot-dashed lines in Figs.~\ref{fig:light-on-shell} and~\ref{fig:A'-invisible}.

\subsection{Converted Mono-Photon Search}
\label{sec:Belle-II-converted-projection}

A small fraction of photons convert to $e^+ e^-$ pairs in the inner detector (see e.g.~\cite{Lees:2011mx}).
While the rate of these events is significantly lower than for non-converted photons, they do allow for significantly better pointing and energy resolution.
The combination of the lower rate and the distinctive nature of the events should make it possible to implement a dedicated trigger for converted mono-photons. 
The improved pointing resolution may 
make it significantly easier to veto mono-photons that are back-to-back with detector regions responsible for photon losses, such as azimuthal gaps in calorimeter coverage. This 
would reduce the background from $\gamma \slashed \gamma$ events, and 
improve the reach around the peak at $m_{\chi \overline \chi}^2 = 0$ compared to what \babar\ achieved in  
the current analysis~\cite{Aubert:2008as}, or even compared to our ``improved \babar'' projection, which already assumed a factor of 10 reduction in the $\gamma \slashed \gamma$ background over~\cite{Aubert:2008as}. 
Moreover, away from the $\gamma \slashed \gamma$ peak, the improved energy resolution may increase  
the power of a bump hunt 
 (although this may not compensate for the reduced amount of data).
 Thus  these factors can potentially strengthen   the search for LDM and are worth a dedicated study by the Belle II collaboration.  

We make projections for such a search assuming that 50/ab of data is collected with a converted-mono-photon trigger, over the energy range $3.2\g < E_\gamma^* < 5.5\g$ (\babar 's High-E region). 
We take the fraction of photons that convert in the tracker to be 5\%, and assume the same cut efficiency as for the standard mono-photon search, giving a combined trigger efficiency of 0.85\%. 
We present two scenarios, one assuming a $\gamma \slashed \gamma$ background reduction factor comparable with our baseline 
assumption of the ``improved \babar'' version, the other assuming a further factor of 10 improvement relative to the ``improved \babar'' version, i.e.~a factor of 100 improvement over the \babar\ analysis in~\cite{Aubert:2008as}.  In both cases, we assume that the continuum background rate is unchanged. These two scenarios are labeled as ``(a)'' and ``(b)'' in Fig.~\ref{fig:light-on-shell} and \ref{fig:A'-invisible}.
We take the energy resolution to be a factor of 2 better than for non-converted photons. Extension to lower $E_{\gamma}$ should also be possible without prescaling given the low conversion fraction (but we do not consider this further). 
An additional 
  improvement of the photon energy resolution by up to a factor of $\sim 2$ may be possible~\cite{Lees:2011mx} due to the improved momentum resolution of low $p_{T}$ tracks.

We show the results with the blue dashed curves in Figs.~\ref{fig:light-on-shell} and~\ref{fig:A'-invisible}. The improvement over current bounds is potentially very substantial, but clearly a dedicated analysis by the Belle II collaboration is required.

%%%%%%%%%%%%%%%%%%%%%%%%%%%%%%%%%%%%%%%%%%%%%%%%%%%%%%
\section{Conclusion}
\label{sec:conclusion}
%%%%%%%%%%%%%%%%%%%%%%%%%%%%%%%%%%%%%%%%%%%%%%%%%%%%%%

Light Dark Matter, coupled to the Standard Model through a light mediator, offers an attractive alternative to the WIMP paradigm. 
However, the parameter space of LDM remains largely unexplored. 
With their large integrated luminosities, current and future 
low-energy $e^+e^-$ colliders offer a uniquely powerful window into LDM parameter space.  

We constrained LDM parameter space using an existing \babar\ search~\cite{Aubert:2008as}.  
We compared this to constraints from 
direct detection experiments and LEP, and in the case of a hidden-photon mediator, also from rare Kaon decays, proton beam dumps, 
 supernova cooling, and QED precision measurements.

Mono-photon searches at future high-luminosity $e^+ e^-$ colliders, such as Belle II, can potentially provide an even more 
powerful probe of LDM and light mediators. 
The most crucial requirement is the implementation of a mono-photon trigger. 
In searching for invisible on-shell mediators, it is important that this be applied for as much run-time as possible 
 (preferably for the entire experiment). 
 For searches that are currently background-limited, the identification of suitable control regions is necessary to estimate the various backgrounds that cannot be computed theoretically in a reliable way, together with the collection of sufficient statistics in such control regions.
In fact, if background estimates  %subtraction 
can be performed with small uncertainties, significant improvements over existing bounds are also possible in the off-shell mediator regions, even with a fraction of the total Belle II mono-photon data.
Additionally, a study of converted mono-photons, using a dedicated trigger, could be extremely powerful. 

Low-energy $e^+e^-$ colliders are one of the most effective probes for light dark matter and light mediators.

%%%%%%%%%%%%%%%%%%%%%%%%%%%%%%%%%%%%%%%%%%%%%%%%%%%%%

\vskip 0.2mm
\begin{center} 

{\bf Note added}
\end{center}

While this paper was in completion,~\cite{Izaguirre:2013uxa,Diamond:2013oda} appeared, 
which focuses on electron beam-dumps, but also 
has some overlap with the \babar\ constraints derived here.  
Past electron-beam dump experiments, like the SLAC milli-charge experiment~\cite{Prinz:1998ua} and E137~\cite{Bjorken:1988as}, 
have sensitivity to LDM, in addition to new experiments at other facilities.  

%%%%%%%%%%%%%%%%%%%%%%%%%%%%%%%%%%%%%%%%%%%%%%%%%%%%%

\vskip 0.2mm
\begin{center} 

{\bf Acknowledgements}
\end{center}
\vskip -1mm
The authors would like to thank Brian Batell, Douglas Bryman, Avner Soffer, Ze'ev Surujon, and especially 
Bertrand Echenard and Yury Kolomensky for useful discussions or correspondence.
RE is supported in part by the DoE Early Career research program DESC0008061 and by a Sloan Foundation Research Fellowship.  
J.M.~is supported by a Simons Postdoctoral Fellowship. 
T.V. is supported in part by a grant from the Israel Science Foundation, the US-Israel Binational Science Foundation,  the EU-FP7 Marie Curie, CIG fellowship and  by the I-CORE Program of the Planning and Budgeting Committee and The Israel Science Foundation (grant NO 1937/12).

%%%%%%%%%%%%%%%%%%%%%%%%%%%%%%%%%%%%%%%%%%%%%%%%%%%%%

\appendix
%\section{Constraints and B-factory Prospects on Invisibly and Visibly Decaying Hidden Photons}
\section{B-factory Prospects on Additional Hidden Photon Scenarios}
\label{sec:invisible-A'}

\begin{figure*}[t!]
\begin{center}
%\fbox{
\includegraphics[width=0.48 \textwidth]{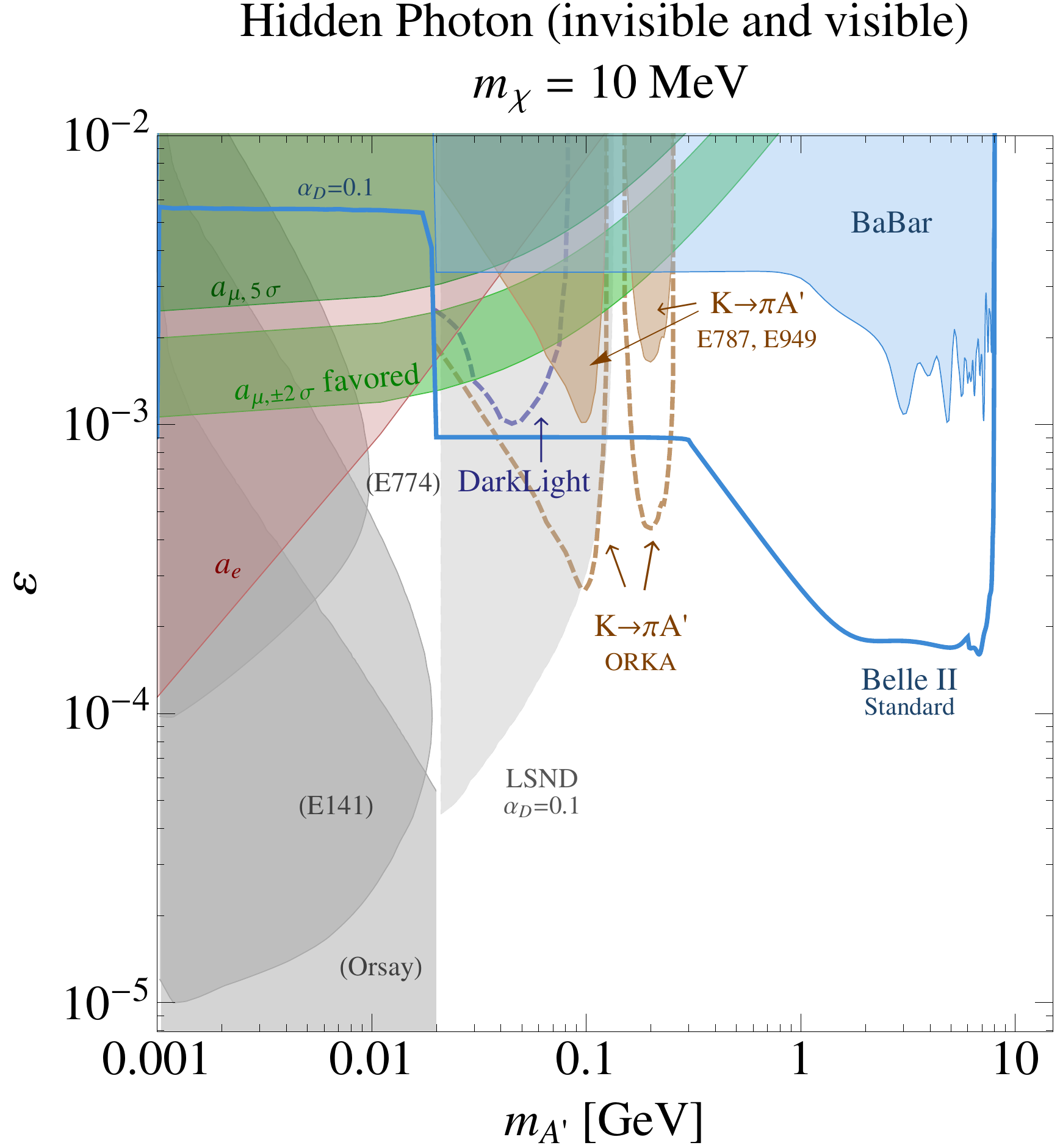} ~~~
\includegraphics[width=0.48 \textwidth]{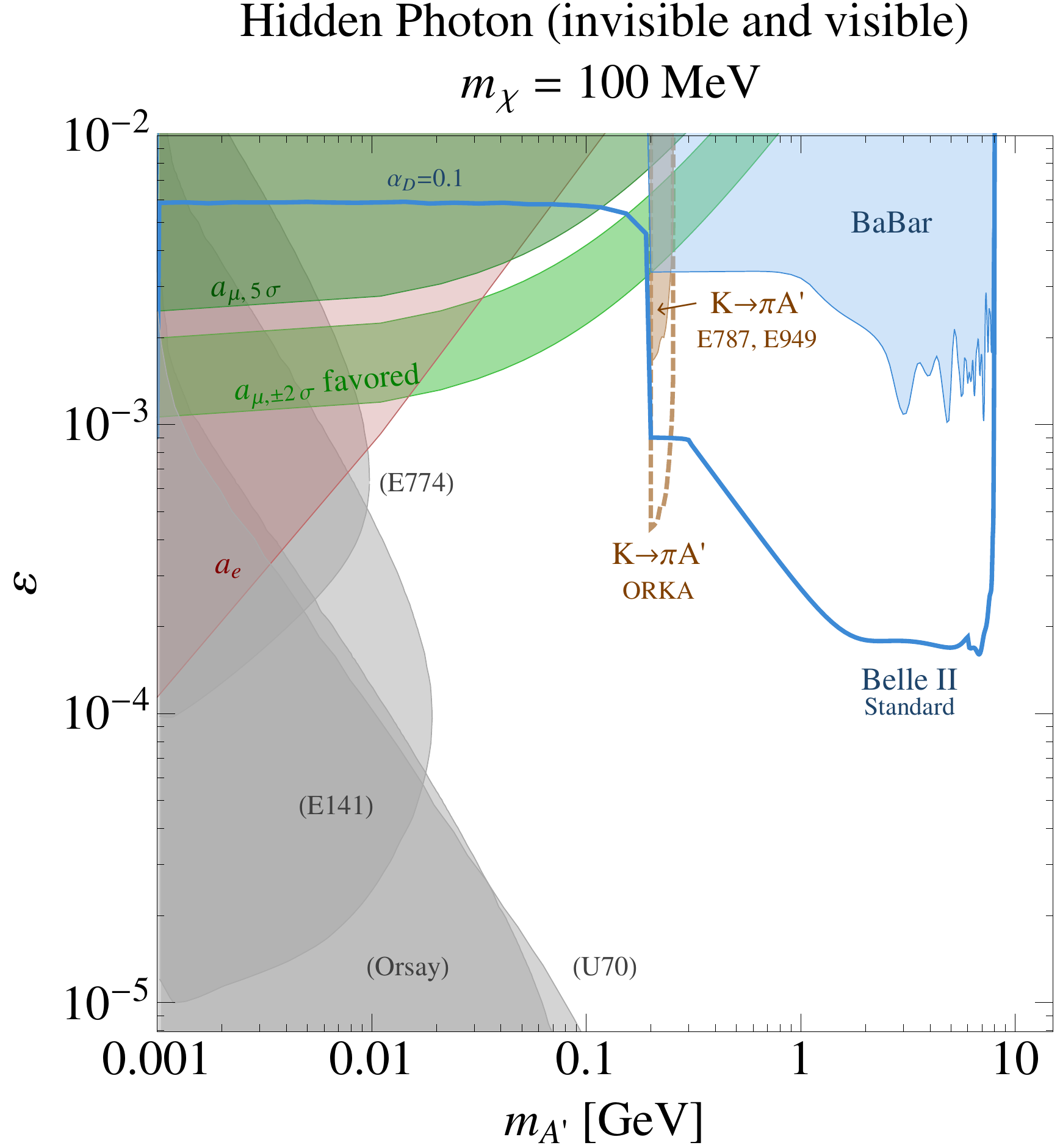}
%}
\caption{
Constraints in the $\varepsilon$ versus $m_{A'}$ plane for {\bf invisibly-decaying} hidden photons assuming they can 
decay either invisibly to a hidden-sector state $\chi$ or visibly to SM matter for $m_{A'} < 2 m_\chi$.  
We show the constraints for fixed mass $m_\chi = 10$~MeV ({\bf left}) or 100~MeV ({\bf right}).  
The bounds from the \babar\ mono-photon data are shown by the blue shaded region. 
Projections for a possible Belle II search is shown with a solid blue line, corresponding to the ``standard'' mono-photon search 
discussed in Sec.~\ref{sec:Belle-II-projections}.  
Various other constraints (shaded regions) and projected sensitivities (dashed lines) are also shown:
the anomalous magnetic moment of the electron ($a_e$, red) and muon ($a_\mu$, blue),
rare kaon decays (brown), 
the upcoming electron fixed-target experiment DarkLight (light blue; shown when kinematically relevant), and LSND 
(light gray; assuming $\alpha_D=0.1$ and that there are no other light hidden-sector states that $\chi$ decays to,  which do not 
interact with the hidden photon).   
In the green shaded region an $A'$ could explain the discrepancy between the measured and predicted SM value of $a_\mu$. 
For $m_{A'} < 2 m_\chi$, we show with gray shaded regions the constraints from visible searches (E141, E774, Orsay, U70 etc.) 
that apply unless there are other decay modes (besides $A' \to \bar \chi \chi$) available for the $A'$. ÊWe do not show the experimental prospects in this case of visible decays. 
More details and references are given in Sec.~\ref{sec:hidden-photon-constraints}.
}
\label{fig:A'-invisible-10-100-MeV}
\end{center}
\end{figure*}

In this short appendix, we show the constraints on invisibly decaying hidden photons, assuming they can decay either invisibly to a 
hidden-sector state $\chi$ or visibly to SM matter for $m_{A'} < 2 m_\chi$.  
While Fig.~\ref{fig:A'-invisible} showed the constraints assuming $m_{A'} < 2 m_\chi$, Fig.~\ref{fig:A'-invisible-10-100-MeV} shows 
the constraints for a fixed LDM mass, $m_\chi = 10$~MeV (left) or 100~MeV (right).  Note that the LSND limit, taken 
from~\cite{deNiverville:2011it}, weakens significantly for larger $m_\chi$ and disappear completely for $m_{\chi}>m_{\pi^0}/2$.  
The limit from rare Kaon decays $K^\pm \to \pi^\pm A'$ disappears when $m_{A'} > 2 m_\chi$.  The other constraints are as 
discussed in Sec.~\ref{sec:hidden-photon-constraints}.  
We do not show the prospects of the experiments that 
have been proposed to search for visible $A'$ decays, but see e.g.~\cite{Hewett:2012ns}.

%%%%%%%%%%%%%%%%%%%%%%%%%%%%%%%%%%%%%%%%%%%%%%%%%%%%%

%\clearpage

\bibliography{ldm-bfactory.bib}

\begin{thebibliography}{100}

\bibitem{Goodman:2010ku}
J.~Goodman {\em et~al.},
\newblock Phys.Rev. {\bf D82}, 116010 (2010), 1008.1783.

\bibitem{Beltran:2010ww}
M.~Beltran, D.~Hooper, E.~W. Kolb, Z.~A. Krusberg, and T.~M. Tait,
\newblock JHEP {\bf 1009}, 037 (2010), 1002.4137.

\bibitem{Bai:2010hh}
Y.~Bai, P.~J. Fox, and R.~Harnik,
\newblock JHEP {\bf 1012}, 048 (2010), 1005.3797.

\bibitem{Goodman:2010yf}
J.~Goodman {\em et~al.},
\newblock Phys.Lett. {\bf B695}, 185 (2011), 1005.1286.

\bibitem{Fox:2011fx}
P.~J. Fox, R.~Harnik, J.~Kopp, and Y.~Tsai,
\newblock Phys.Rev. {\bf D84}, 014028 (2011), 1103.0240.

\bibitem{Rajaraman:2011wf}
A.~Rajaraman, W.~Shepherd, T.~M. Tait, and A.~M. Wijangco,
\newblock Phys.Rev. {\bf D84}, 095013 (2011), 1108.1196.

\bibitem{Fox:2011pm}
P.~J. Fox, R.~Harnik, J.~Kopp, and Y.~Tsai,
\newblock Phys.Rev. {\bf D85}, 056011 (2012), 1109.4398.

\bibitem{Fox:2012ee}
P.~J. Fox, R.~Harnik, R.~Primulando, and C.-T. Yu,
\newblock Phys.Rev. {\bf D86}, 015010 (2012), 1203.1662.

\bibitem{Chatrchyan:2012tea}
CMS Collaboration, S.~Chatrchyan {\em et~al.},
\newblock Phys.Rev.Lett. {\bf 108}, 261803 (2012), 1204.0821.

\bibitem{Aad:2012fw}
ATLAS Collaboration, G.~Aad {\em et~al.},
\newblock Phys.Rev.Lett. {\bf 110}, 011802 (2013), 1209.4625.

\bibitem{An:2012ue}
H.~An, R.~Huo, and L.-T. Wang,
\newblock Phys.Dark Univ. {\bf 2}, 50 (2013), 1212.2221.

\bibitem{Fox:2012ru}
P.~J. Fox and C.~Williams,
\newblock Phys.Rev. {\bf D87}, 054030 (2013), 1211.6390.

\bibitem{Zhou:2013fla}
N.~Zhou, D.~Berge, and D.~Whiteson,
\newblock (2013), 1302.3619.

\bibitem{Baltz:2006fm}
E.~A. Baltz, M.~Battaglia, M.~E. Peskin, and T.~Wizansky,
\newblock Phys.Rev. {\bf D74}, 103521 (2006), hep-ph/0602187.

\bibitem{Dreiner:2006sb}
H.~K. Dreiner, O.~Kittel, and U.~Langenfeld,
\newblock Phys.Rev. {\bf D74}, 115010 (2006), hep-ph/0610020.

\bibitem{Dreiner:2007vm}
H.~K. Dreiner, O.~Kittel, and U.~Langenfeld,
\newblock Eur.Phys.J. {\bf C54}, 277 (2008), hep-ph/0703009.

\bibitem{Dreiner:2009ic}
H.~K. Dreiner {\em et~al.},
\newblock Eur.Phys.J. {\bf C62}, 547 (2009), 0901.3485.

\bibitem{Bird:2004ts}
C.~Bird, P.~Jackson, R.~V. Kowalewski, and M.~Pospelov,
\newblock Phys.Rev.Lett. {\bf 93}, 201803 (2004), hep-ph/0401195.

\bibitem{McElrath:2005bp}
B.~McElrath,
\newblock Phys.Rev. {\bf D72}, 103508 (2005), hep-ph/0506151.

\bibitem{Fayet:2006sp}
P.~Fayet,
\newblock Phys.Rev. {\bf D74}, 054034 (2006), hep-ph/0607318.

\bibitem{Bird:2006jd}
C.~Bird, R.~V. Kowalewski, and M.~Pospelov,
\newblock Mod.Phys.Lett. {\bf A21}, 457 (2006), hep-ph/0601090.

\bibitem{Rubin:2006gc}
CLEO Collaboration, P.~Rubin {\em et~al.},
\newblock Phys.Rev. {\bf D75}, 031104 (2007), hep-ex/0612051.

\bibitem{Tajima:2006nc}
Belle Collaboration, O.~Tajima {\em et~al.},
\newblock Phys.Rev.Lett. {\bf 98}, 132001 (2007), hep-ex/0611041.

\bibitem{Kahn:2007ru}
Y.~Kahn, M.~Schmitt, and T.~M. Tait,
\newblock Phys.Rev. {\bf D78}, 115002 (2008), 0712.0007.

\bibitem{Dorokhov:2007bd}
A.~E. Dorokhov and M.~A. Ivanov,
\newblock Phys.Rev. {\bf D75}, 114007 (2007), 0704.3498.

\bibitem{Fayet:2007ua}
P.~Fayet,
\newblock Phys.Rev. {\bf D75}, 115017 (2007), hep-ph/0702176.

\bibitem{Fayet:2009tv}
P.~Fayet,
\newblock Phys.Rev. {\bf D81}, 054025 (2010), 0910.2587.

\bibitem{Yeghiyan:2009xc}
G.~K. Yeghiyan,
\newblock Phys.Rev. {\bf D80}, 115019 (2009), 0909.4919.

\bibitem{Dorokhov:2009xs}
A.~Dorokhov, M.~Ivanov, and S.~Kovalenko,
\newblock Phys.Lett. {\bf B677}, 145 (2009), 0903.4249.

\bibitem{delAmoSanchez:2010ac}
BaBar Collaboration, P.~del Amo~Sanchez {\em et~al.},
\newblock Phys.Rev.Lett. {\bf 107}, 021804 (2011), 1007.4646.

\bibitem{Badin:2010uh}
A.~Badin and A.~A. Petrov,
\newblock Phys.Rev. {\bf D82}, 034005 (2010), 1005.1277.

\bibitem{Echenard:2012iq}
B.~Echenard,
\newblock Mod.Phys.Lett. {\bf A27}, 1230016 (2012), 1205.3505.

\bibitem{Calderini:2012ar}
G.~Calderini,
\newblock EPJ Web Conf. {\bf 28}, 04005 (2012), 1204.4281.

\bibitem{Borodatchenkova:2005ct}
N.~Borodatchenkova, D.~Choudhury, and M.~Drees,
\newblock Phys.Rev.Lett. {\bf 96}, 141802 (2006), hep-ph/0510147.

\bibitem{Essig:2009nc}
R.~Essig, P.~Schuster, and N.~Toro,
\newblock Phys.Rev. {\bf D80}, 015003 (2009), 0903.3941.

\bibitem{Reece:2009un}
M.~Reece and L.-T. Wang,
\newblock JHEP {\bf 0907}, 051 (2009), 0904.1743.

\bibitem{Aubert:2008as}
BaBar Collaboration, B.~Aubert {\em et~al.},
\newblock (2008), 0808.0017.

\bibitem{Boehm:2003hm}
C.~Boehm and P.~Fayet,
\newblock Nucl.Phys. {\bf B683}, 219 (2004), hep-ph/0305261.

\bibitem{Boehm:2003ha}
C.~Boehm, P.~Fayet, and J.~Silk,
\newblock Phys.Rev. {\bf D69}, 101302 (2004), hep-ph/0311143.

\bibitem{Strassler:2006im}
M.~J. Strassler and K.~M. Zurek,
\newblock Phys.Lett. {\bf B651}, 374 (2007), hep-ph/0604261.

\bibitem{Hooper:2008im}
D.~Hooper and K.~M. Zurek,
\newblock Phys.Rev. {\bf D77}, 087302 (2008), 0801.3686.

\bibitem{ArkaniHamed:2008qn}
N.~Arkani-Hamed, D.~P. Finkbeiner, T.~R. Slatyer, and N.~Weiner,
\newblock Phys.Rev. {\bf D79}, 015014 (2009), 0810.0713.

\bibitem{Pospelov:2008jd}
M.~Pospelov and A.~Ritz,
\newblock Phys. Lett. {\bf B671}, 391 (2009).

\bibitem{Pospelov:2007mp}
M.~Pospelov, A.~Ritz, and M.~B. Voloshin,
\newblock Phys.Lett. {\bf B662}, 53 (2008), 0711.4866.

\bibitem{Cholis:2008vb}
I.~Cholis, L.~Goodenough, and N.~Weiner,
\newblock Phys.Rev. {\bf D79}, 123505 (2009), 0802.2922.

\bibitem{Essig:2010ye}
R.~Essig, J.~Kaplan, P.~Schuster, and N.~Toro,
\newblock Submitted to Physical Review D  (2010), 1004.0691.

\bibitem{Ruderman:2009tj}
J.~T. Ruderman and T.~Volansky,
\newblock JHEP {\bf 1002}, 024 (2010), 0908.1570.

\bibitem{Falkowski:2010cm}
A.~Falkowski, J.~T. Ruderman, T.~Volansky, and J.~Zupan,
\newblock JHEP {\bf 1005}, 077 (2010), 1002.2952.

\bibitem{Morrissey:2009ur}
D.~E. Morrissey, D.~Poland, and K.~M. Zurek,
\newblock JHEP {\bf 0907}, 050 (2009), 0904.2567.

\bibitem{Nomura:2008ru}
Y.~Nomura and J.~Thaler,
\newblock Phys.Rev. {\bf D79}, 075008 (2009), 0810.5397.

\bibitem{Baumgart:2009tn}
M.~Baumgart, C.~Cheung, J.~T. Ruderman, L.-T. Wang, and I.~Yavin,
\newblock JHEP {\bf 0904}, 014 (2009), 0901.0283.

\bibitem{Feng:2008ya}
J.~L. Feng and J.~Kumar,
\newblock Phys.Rev.Lett. {\bf 101}, 231301 (2008), 0803.4196.

\bibitem{Abel:2008ai}
S.~Abel, M.~Goodsell, J.~Jaeckel, V.~Khoze, and A.~Ringwald,
\newblock JHEP {\bf 0807}, 124 (2008), 0803.1449.

\bibitem{Jaeckel:2010ni}
J.~Jaeckel and A.~Ringwald,
\newblock Ann.Rev.Nucl.Part.Sci. {\bf 60}, 405 (2010), 1002.0329.

\bibitem{Cohen:2010kn}
T.~Cohen, D.~J. Phalen, A.~Pierce, and K.~M. Zurek,
\newblock Phys.Rev. {\bf D82}, 056001 (2010), 1005.1655.

\bibitem{Lin:2011gj}
T.~Lin, H.-B. Yu, and K.~M. Zurek,
\newblock Phys.Rev. {\bf D85}, 063503 (2012), 1111.0293.

\bibitem{MarchRussell:2012hi}
J.~March-Russell, J.~Unwin, and S.~M. West,
\newblock JHEP {\bf 1208}, 029 (2012), 1203.4854.

\bibitem{Loeb:2010gj}
A.~Loeb and N.~Weiner,
\newblock Phys.Rev.Lett. {\bf 106}, 171302 (2011), 1011.6374.

\bibitem{Tulin:2013teo}
S.~Tulin, H.-B. Yu, and K.~M. Zurek,
\newblock (2013), 1302.3898.

\bibitem{Holdom:1985ag}
B.~Holdom,
\newblock Phys. Lett. {\bf B166}, 196 (1986).

\bibitem{Dienes:1996zr}
K.~R. Dienes, C.~F. Kolda, and J.~March-Russell,
\newblock Nucl.Phys. {\bf B492}, 104 (1997), hep-ph/9610479.

\bibitem{Galison:1983pa}
P.~Galison and A.~Manohar,
\newblock Phys.Lett. {\bf B136}, 279 (1984).

\bibitem{Bjorken:2009mm}
J.~D. Bjorken, R.~Essig, P.~Schuster, and N.~Toro,
\newblock Phys. Rev. {\bf D80}, 075018 (2009).

\bibitem{Hewett:2012ns}
J.~Hewett {\em et~al.},
\newblock (2012), 1205.2671.

\bibitem{Hook:2010tw}
A.~Hook, E.~Izaguirre, and J.~G. Wacker,
\newblock Adv.High Energy Phys. {\bf 2011}, 859762 (2011), 1006.0973.

\bibitem{Lees:2013rw}
BaBar Collaboration, J.~Lees {\em et~al.},
\newblock (2013), 1301.2703.

\bibitem{Alwall:2011uj}
J.~Alwall, M.~Herquet, F.~Maltoni, O.~Mattelaer, and T.~Stelzer,
\newblock JHEP {\bf 1106}, 128 (2011), 1106.0522.

\bibitem{Archilli:2011zc}
F.~Archilli {\em et~al.},
\newblock Phys.Lett. {\bf B706}, 251 (2012), 1110.0411.

\bibitem{Ablikim:2007ek}
BES Collaboration, M.~Ablikim {\em et~al.},
\newblock Phys.Rev.Lett. {\bf 100}, 192001 (2008), 0710.0039.

\bibitem{Dharmapalan:2012xp}
MiniBooNE Collaboration, R.~Dharmapalan {\em et~al.},
\newblock (2012), 1211.2258.

\bibitem{Adler:2004hp}
E787 Collaboration, S.~Adler {\em et~al.},
\newblock Phys.Rev. {\bf D70}, 037102 (2004), hep-ex/0403034.

\bibitem{Artamonov:2008qb}
E949 Collaboration, A.~Artamonov {\em et~al.},
\newblock Phys.Rev.Lett. {\bf 101}, 191802 (2008), 0808.2459.

\bibitem{Brod:2010hi}
J.~Brod, M.~Gorbahn, and E.~Stamou,
\newblock Phys.Rev. {\bf D83}, 034030 (2011), 1009.0947.

\bibitem{deNiverville:2011it}
P.~deNiverville, M.~Pospelov, and A.~Ritz,
\newblock Phys.Rev. {\bf D84}, 075020 (2011), 1107.4580.

\bibitem{Artamonov:2009sz}
BNL-E949 Collaboration, A.~Artamonov {\em et~al.},
\newblock Phys.Rev. {\bf D79}, 092004 (2009), 0903.0030.

\bibitem{Pospelov:2008zw}
M.~Pospelov,
\newblock Phys.Rev. {\bf D80}, 095002 (2009), 0811.1030.

\bibitem{DAmbrosio:1998yj}
G.~D'Ambrosio, G.~Ecker, G.~Isidori, and J.~Portoles,
\newblock JHEP {\bf 9808}, 004 (1998), hep-ph/9808289.

\bibitem{E.T.WorcesterfortheORKA:2013cya}
E. T. Worcester for the ORKA collaboration,
\newblock (2013), 1305.7245.

\bibitem{ORKA-talk}
ORKA Collaboration, D.~Bryman,
\newblock {ORKA at Fermilab: Seeking New Physics with Measurements of the
  ``Golden Kaon'' Decay $K^+ \to \pi^+ \bar\nu\nu$},
\newblock talk given at Argonne Intensity Frontier Workshop, April 2013.

\bibitem{NA62}
NA62,
\newblock {2013 NA62 status report to the CERN SPSC}.

\bibitem{Davier:2010nc}
M.~Davier, A.~Hoecker, B.~Malaescu, and Z.~Zhang,
\newblock Eur.Phys.J. {\bf C71}, 1515 (2011), 1010.4180.

\bibitem{Bennett:2006fi}
Muon G-2 Collaboration, G.~Bennett {\em et~al.},
\newblock Phys.Rev. {\bf D73}, 072003 (2006), hep-ex/0602035.

\bibitem{Aoyama:2012wj}
T.~Aoyama, M.~Hayakawa, T.~Kinoshita, and M.~Nio,
\newblock Phys.Rev.Lett. {\bf 109}, 111807 (2012), 1205.5368.

\bibitem{2011PhRvL.106h0801B}
R.~{Bouchendira}, P.~{Clad{\'e}}, S.~{Guellati-Kh{\'e}lifa}, F.~{Nez}, and
  F.~{Biraben},
\newblock Physical Review Letters {\bf 106}, 080801 (2011), 1012.3627.

\bibitem{2008PhRvL.100l0801H}
D.~{Hanneke}, S.~{Fogwell}, and G.~{Gabrielse},
\newblock Physical Review Letters {\bf 100}, 120801 (2008), 0801.1134.

\bibitem{Davoudiasl:2012ig}
H.~Davoudiasl, H.-S. Lee, and W.~J. Marciano,
\newblock Phys.Rev. {\bf D86}, 095009 (2012), 1208.2973.

\bibitem{Endo:2012hp}
M.~Endo, K.~Hamaguchi, and G.~Mishima,
\newblock Phys.Rev. {\bf D86}, 095029 (2012), 1209.2558.

\bibitem{Batell:2009yf}
B.~Batell, M.~Pospelov, and A.~Ritz,
\newblock Phys. Rev. {\bf D79}, 115008 (2009).

\bibitem{Freytsis:2009bh}
M.~Freytsis, G.~Ovanesyan, and J.~Thaler,
\newblock JHEP {\bf 1001}, 111 (2010), 0909.2862.

\bibitem{Batell:2009di}
B.~Batell, M.~Pospelov, and A.~Ritz,
\newblock Phys.Rev. {\bf D80}, 095024 (2009), 0906.5614.

\bibitem{:2009pw}
BABAR, B.~Aubert {\em et~al.},
\newblock (2009), arXiv:0908.2821.

\bibitem{Bossi:2009uw}
F.~Bossi,
\newblock (2009), 0904.3815.

\bibitem{Essig:2010xa}
R.~Essig, P.~Schuster, N.~Toro, and B.~Wojtsekhowski,
\newblock JHEP {\bf 02}, 009 (2011).

\bibitem{Essig:2010gu}
R.~Essig, R.~Harnik, J.~Kaplan, and N.~Toro,
\newblock Phys. Rev. {\bf D82}, 113008 (2010).

\bibitem{HPS}
{e.g.~{\small{{https://confluence.slac.stanford.edu/display/hpsg/}}}.}

\bibitem{Abrahamyan:2011gv}
APEX, S.~Abrahamyan {\em et~al.},
\newblock Phys. Rev. Lett. {\bf 107}, 191804 (2011), 1108.2750.

\bibitem{Merkel:2011ze}
A1, H.~Merkel {\em et~al.},
\newblock Phys. Rev. Lett. {\bf 106}, 251802 (2011).

\bibitem{Archilli:2011nh}
F.~Archilli {\em et~al.},
\newblock (2011), arXiv:1107.2531.

\bibitem{DarkLight}
DarkLight Collaboration, G.~Bennett {\em et~al.}

\bibitem{Wojtsekhowski:2012zq}
B.~Wojtsekhowski, D.~Nikolenko, and I.~Rachek,
\newblock (2012), 1207.5089.

\bibitem{Kahn:2012br}
Y.~Kahn and J.~Thaler,
\newblock Phys.Rev. {\bf D86}, 115012 (2012), 1209.0777.

\bibitem{deNiverville:2012ij}
P.~deNiverville, D.~McKeen, and A.~Ritz,
\newblock Phys.Rev. {\bf D86}, 035022 (2012), 1205.3499.

\bibitem{Dent:2012mx}
J.~B. Dent, F.~Ferrer, and L.~M. Krauss,
\newblock (2012), 1201.2683.

\bibitem{Davidson:2000hf}
S.~Davidson, S.~Hannestad, and G.~Raffelt,
\newblock JHEP {\bf 0005}, 003 (2000), hep-ph/0001179.

\bibitem{Dreiner:SN}
H.~K. Dreiner, J.-F. Fortin, C.~Hanhart, and L.~Ubaldi,
\newblock (to appear).

\bibitem{Dreiner:2013tja}
H.~K. Dreiner, J.-F. Fortin, J.~Isern, and L.~Ubaldi,
\newblock Phys.Rev. {\bf D88}, 043517 (2013), 1303.7232.

\bibitem{Altmann:2001ax}
M.~F. Altmann {\em et~al.},
\newblock (2001), astro-ph/0106314.

\bibitem{Barreto:2011zu}
DAMIC Collaboration, J.~Barreto {\em et~al.},
\newblock Phys.Lett. {\bf B711}, 264 (2012), 1105.5191.

\bibitem{Essig:2012yx}
R.~Essig, A.~Manalaysay, J.~Mardon, P.~Sorensen, and T.~Volansky,
\newblock Phys.Rev.Lett. {\bf 109}, 021301 (2012), 1206.2644.

\bibitem{Essig:2011nj}
R.~Essig, J.~Mardon, and T.~Volansky,
\newblock Phys.Rev. {\bf D85}, 076007 (2012), 1108.5383.

\bibitem{Abe:2010sj}
Belle II Collaboration, T.~Abe,
\newblock (2010), 1011.0352.

\bibitem{Belle-II-talk:2013}
Belle II Collaboration, L.~Piilonen,
\newblock {Status and Prospects of SuperKEKB/Belle II},
\newblock talk given at BNL Forum 2013.

\bibitem{Lees:2011mx}
BaBar Collaboration, J.~Lees {\em et~al.},
\newblock Phys.Rev. {\bf D84}, 072002 (2011), 1104.5254.

\bibitem{Izaguirre:2013uxa}
E.~Izaguirre, G.~Krnjaic, P.~Schuster, and N.~Toro,
\newblock (2013), 1307.6554.

\bibitem{Diamond:2013oda}
M.~D. Diamond and P.~Schuster,
\newblock (2013), 1307.6861.

\bibitem{Prinz:1998ua}
A.~Prinz {\em et~al.},
\newblock Phys.Rev.Lett. {\bf 81}, 1175 (1998), hep-ex/9804008.

\bibitem{Bjorken:1988as}
J.~Bjorken {\em et~al.},
\newblock Phys.Rev. {\bf D38}, 3375 (1988).

\end{thebibliography}
\bibliographystyle{h-physrev.bst}

\end{document}